\documentclass[aps,twocolumn]{revtex4-1}
\usepackage{amsmath,amssymb}
\usepackage{slashed}
\usepackage{graphicx}
\usepackage{bm}
\usepackage[T1]{fontenc}
\usepackage[utf8]{inputenc}
\usepackage{gauss} 
\usepackage[top=1.0in,bottom=1.0in,left=1.0in,right=1.0in]{geometry}
\usepackage[colorlinks,linkcolor=blue,citecolor=blue]{hyperref}
\newcommand{\be}{\begin{equation}}
\newcommand{\ee}{\end{equation}}
\newcommand{\bea}{\begin{eqnarray}}
\newcommand{\eea}{\end{eqnarray}}
\newcommand{\ba}{\begin{eqnarray}}
\newcommand{\ea}{\end{eqnarray}}

\begin{document}

\title{Hadronic structure on the light-front  IV.\\
Heavy and light Baryons}

\author{Edward Shuryak}
\email{edward.shuryak@stonybrook.edu}
\affiliation{Center for Nuclear Theory, Department of Physics and Astronomy, Stony Brook University, Stony Brook, New York 11794--3800, USA}

\author{Ismail Zahed}
\email{ismail.zahed@stonybrook.edu}
\affiliation{Center for Nuclear Theory, Department of Physics and Astronomy, Stony Brook University, Stony Brook, New York 11794--3800, USA}

\begin{abstract}
This  work is a continuation in our  series of papers,  that addresses quark models of 
hadronic structure
 on the light front. The chief focus of this paper is the quantum-mechanical solution of 
 the three-quark model Hamiltonian,  describing baryons. In  Jacobi coordinates,  we
use a harmonic oscillator basis for the transverse directions. For the longitudinal momentum fractions
 $x_i$, the pertinent basis follows from   quantum mechanics in a ``triangular cup" potential, which we solve exactly.
 We calculate the masses and light front wave functions for the flavor symmetric $\frac 32^+$ baryons $bbb,ccc,sss,uuu$.  
 \end{abstract}

\maketitle
\section{Introduction} 
Since this is the fourth paper of the series \cite{Shuryak:2021fsu,Shuryak:2021hng,Shuryak:2021mlh}, it does not need 
an extensive introduction. Its main goal is to
bridge the gaps between subfields of hadronic physics, with our general direction 
being from (I) {\em the vacuum structure} in its Euclidean formulation (instantons and lattice),
to (ii) {\em the hadronic structure} and quark-quark interactions and resulting spectroscopy, to
(iii) {\em the hadronic structure on the  light front} with its novel Hamiltonians and wave functions.

The connection between (i) and (ii) is provided by nonlocal gauge field correlators, such as 
e.g. correlations of Wilson lines defining static quark potentials. Using lattice or semiclassical
models of the vacuum fields, one can evaluate them. The connection between (ii) and (iii) 
is  less developed, as neither spectroscopists nor people studying partonic observables
were inclined to study them. (The former community is now living through a deluge of new hadrons 
discovered lately, and is rather busy.) So, let us emphasize some of the reasons  for its
development.

Standard spectroscopy  (in the center of mass or CM frame)
uses rather different tools for states made of heavy and light quarks. 
The differences stem from both kinematical and dynamical reasons. 

Kinematically, the heavy quarkonia
can be treated nonrelativistically, using the Schroedinger equation and perturbative effective theories like pHQCD, 
while the light quarks are studied with relativistic tools such as the Bethe-Salpeter equation and
the like. 
(In fact, even the standard  approaches to heavy quarkonia are not so accurate, as
one might get from textbooks. Say,
for charm quark,  the typical velocity is not really small, $v\sim \frac 12$ or so.)

Dynamically, there are important differences between heavy and light quark interactions. 
Indeed, light quark physics is tightly bound to the issue of chiral symmetry breaking,
and its root causes -- strong short-range effects described by NJL operators
or instanton-based t'Hooft Lagrangian. Most of that was well understood in the 1990's 
and need not be repeated here.

However,
 as we have shown in \cite{Shuryak:2021fsu}, a dilute instanton ensemble  is only one part
 of the vacuum fluctuations related with gauge topology at low resolution, and when one studies gauge field
 observables one finds larger effects at moderatly higher resolution. Even for heavy quarkonia, we argued that a
 ``dense vacuum" with instanton-antiinstanton pairs (incomplete tunneling through a topologicl barrier) 
 contributes to Wilson line correlators, with and without magnetic fields, and
 generates a good fraction of the central and spin-dependent forces. This raises a question 
 of how one can include
 those effects for light quarks.
 
Fortunately, both these kinematical and dynamical issues  are much less severe
on the light front. The kinematics in this case is simply relativistic for all masses. There are no
 sudden changes, as one go from heavy to light quarks. Quark masses enter the $H_{LF}$ in a very uniform way and (as
we have shown in the previous papers of the series \cite{Shuryak:2021fsu,Shuryak:2021hng,Shuryak:2021mlh}), one can consistently derive the 
mesonic properties from $\bar b b$ to light $\bar q q$ by the same tools. Indeed, in 
the first approximation, the transverse oscillator Hamiltonian generates near-linear Regge 
dependences of $M^2$,  on the  principal quantum number $n$ and angular momentum $m$.
Dynamical issues also get less severe. In particular, on the  light front,  even light quarks 
can be ``eikonalized"  as they move along approximately straight lines.

\subsection{Single-flavor baryons}
Baryons are just another application  of 
the tools developed along the lines mentioned above, but this time for three-quark systems. There are important  technical issues here as well, 
as the barrier between ``relative motion" in mesons and baryons, is   due to the differences between the obvious variables
describing the relative motion of two particles,  and the nontrivial choices of variable for few-body
quantum mechanics. We will address those below, but before that let us add some general
remarks.

In principle,  another (non-technical) issue is related with the so called ``color junction" of three strings. 
The quadratic confining potential of a  ``star"  (or $Y$) model fixes the junction at  the origin,  with no dynamics.
For static potentials we can probe the effects of the junction by changing its location. However,  the junction
is in general dynamical, and should be treated as a {\em fourth body}.  In general, the effective string Lagrangians
carry also  boundary terms, and a junction-line connecting the three world-volumes should also be added as a boundary contribution.
The dynamics of the junction can only be ignored if it is heavy, but in so far there is no empirical indication  of that.
This problem remains to our knowledge open.

This notwithstanding, one should note that in the last decade,
we have seen  discoveries of multiple new hadrons in the so-called heavy-light 
sector, including $QQq$ baryons and tetraquarks of the type $\bar Q Q \bar q q$ and $QQ \bar q \bar q$.
 Calculations for similar states with five and six quarks are ongoing by many groups.
They will shed more light on  the issue of quark-quark interactions.
Also, baryons too have a 5-quark sector, responsible for the {\em antiquark sea}, well studied 
experimentally in the case of the proton and neutron. Their flavor structure has been recently
discussed by one of us \cite{Shuryak:2019zhv}. 

Non-relativistic and semi-relativisitc constituent quark models, have been developed since the 1960's, and they exist
in numerous versions. One well documented (and still widely used as a reference point) approach is that
by Isgur and Karl \cite{Isgur:1979be}, which was updated for heavy quark states, see e.g. \cite{0711.2492}.
These authors treated confinement by an oscillatory
potential, which  methodically will turn out to be similar to our $H_{LF}$ (but for squared mass, not energy).  A well known problem with
the model,  is its predictions of many more baryonic states than what was experimentally  observed.

The focus of this paper is on
on basic baryons which are completely symmetric in flavor, 
such as $\Delta^{++}_{uuu}, \Omega_{sss}^-,\Omega^{++}_{ccc},\Omega_{bbb}^-$. Yes, although only the first two of them have been observed. (According to estimates, $\Omega^{++}_{ccc}$ will be discovered in the next LHC run.)
The reason is that flavor asymmetric pairs such as $ud,us,ds...$ have deeply bound diquark correlations
which will be the subject of our next paper.

General considerations for tthese hadrons are well known, e.g. summarized in the early note by Bjorken~\cite{Bjorken:1986xpa}.
If the color part of the wave function is antisymmetric and the flavor part is symmetric,
then Fermi statistics requires the  spin-orbital part to be symmetric as well. The simplest
one, with no orbital motion, then fixes spins to be e.g. $\uparrow \uparrow \uparrow $
and the global quantum number to be $\frac 32^+$. We will  focus on  the
sector with zero orbital momentum, thereby avoiding the inclusion of spin-orbit mixing  
(on which we focused in the previous paper \cite{Shuryak:2021mlh} for mesons).

In Table \ref{tab_1} we show the
quark and baryon masses, as well as the binding of the lowest  $\frac 32^+$ 
states
according to Ref.~\cite{0711.2492}. We note that as we move from heavy to light baryons,
 the binding changes from negative to positive values (for the  sum of the masses). This is  due to the attractive Coulomb
interaction at small distances, whose role dramatically decreases for lighter quarks, as their states become larger in size.

\begin{table}[htp]
\caption{Baryon masses and binding energies ( all in $GeV$) for different quark flavors.
Two baryon masses in the last two rows are experimental, all other numbers are as
used in Ref.~\cite{0711.2492}.}
\begin{center}
\begin{tabular}{|c|c|c|c|c|c|} \hline
 & $m_Q$  &  $M_{QQQ} ^{3/2^+}$ & $M_{QQQ}^{ 3/2^+}-3m_Q$ \\  \hline 
  b & 5.2019 &    14.834  & -0.7717  \\ 
 c & 1.8182&   4.965  & -0.4896  \\ 
 s &  .5553&   1.672  & 0.006 \\ 
 q & .2848 &   1.232  &  0.3776  \\
 \hline
\end{tabular}
\end{center}
\label{tab_1}
\end{table}%
 
 \begin{figure}[t!]
\begin{center}
\includegraphics[width=6cm]{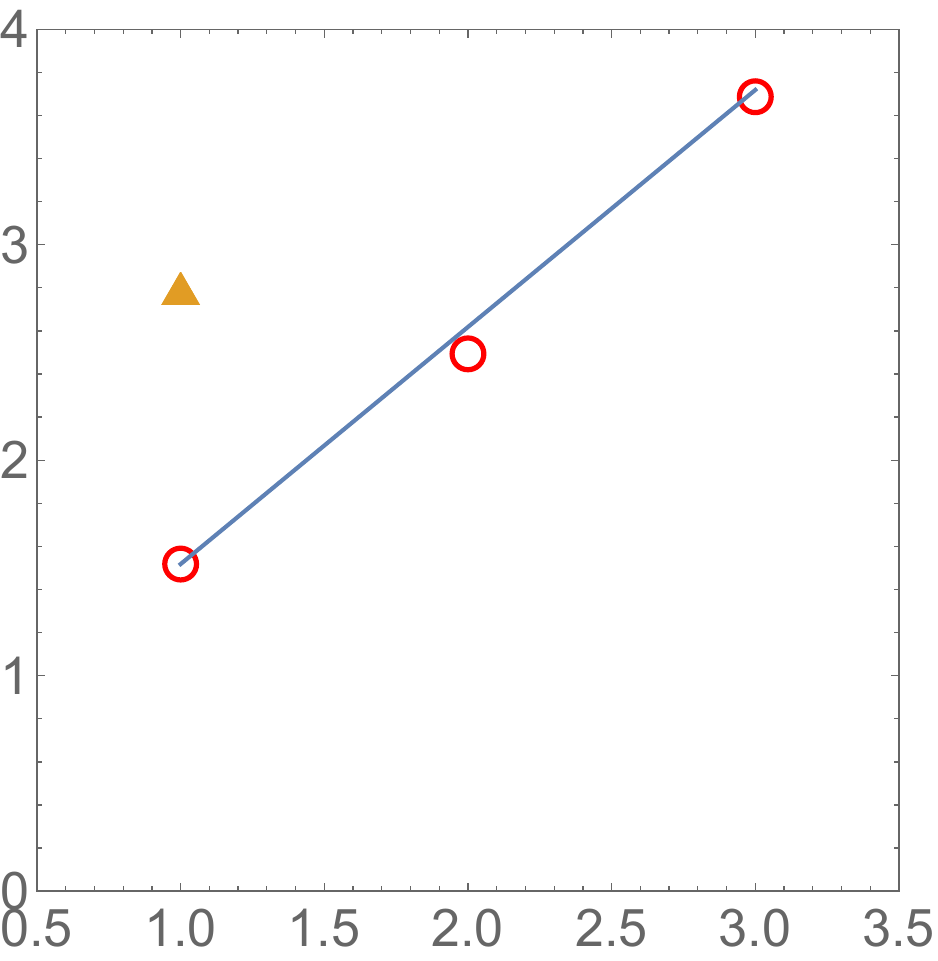}
\includegraphics[width=6cm]{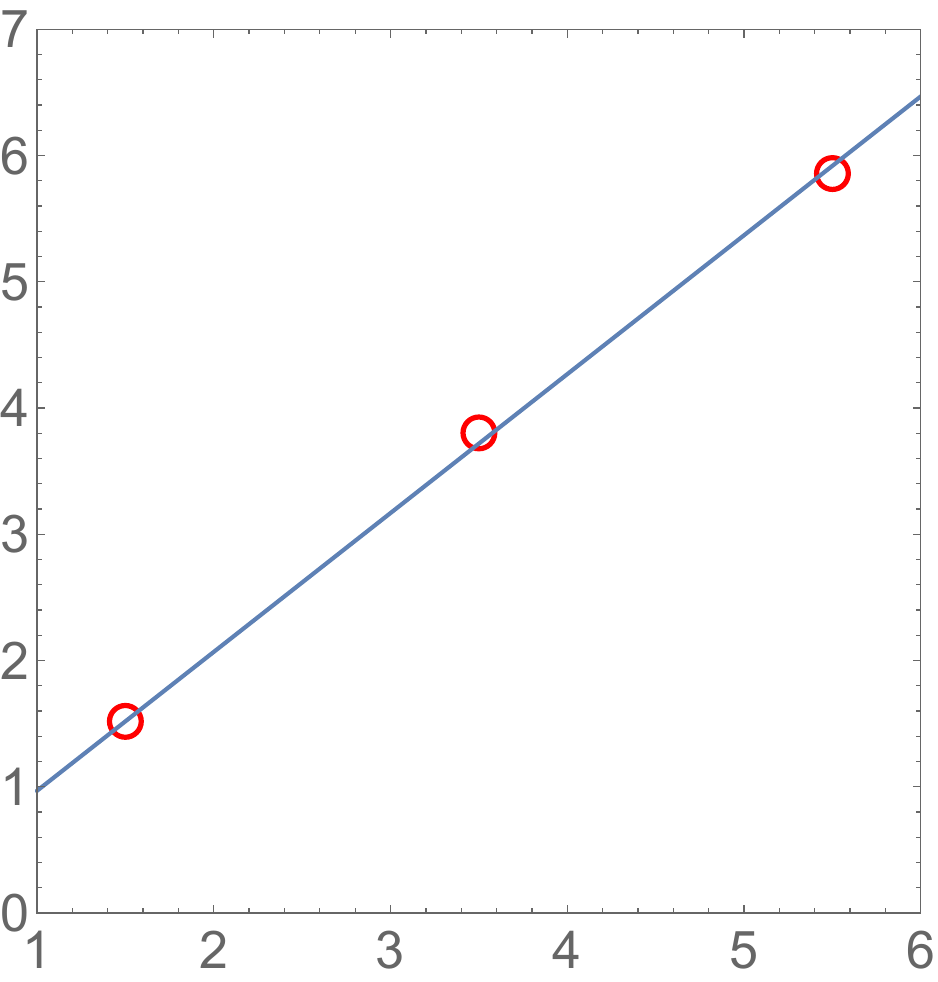}
\caption{Upper plot:  red circles are the  squared masses of Delta resonances  $M^2_\Delta(n+1,3/2)\, (GeV)^2$ from PDG tables
versus the principle quantum number $n+1=1,2,3$. The brown triangle corresponds to  the triple-strange  baryon $M^2_\Omega(1,\frac 32)$.
The lower plot shows the dependence  of  the  Delta resonances  $M^2_\Delta(1,J)$ with angular momenta $J=\frac 32,\frac 72,\frac{11}2$.
Both  straight lines have the same slope $1.1 \, GeV^2$. }
\label{fig_deltas_nj}
\end{center}
\end{figure}
The dependence of the masses and wave functions of these {\em ground state} baryons,  on the quark mass is of course only one issue to be
considered. Another is their spectrum, in particular the dependence on the {\em principle quantum numbers} $n$,
 and {\em total angular momentum} $J$. It is well known that  confining strings lead to specific Regge trajectories,
both for mesons and baryons. For example, we show in Fig.\ref{fig_deltas_nj} that the squared masses of various $\Delta$ resonances
follow linear trajectories, versus  angular  momentum $J$. Two further remarkable observations are: (i) both plots have the same slope; and (ii) this slope is the same as for mesons.  
This  leads to the  well known difficulty of a ``star" (or Y) and other stringy models: they  do not  yield the correct
Regge slope, as the tension of three strings is different from that of a single string in mesons. 
The qualitative resolution of this difficulty for the nucleons  is believed to be a quark-diquark configuration, with a single string between them. However, a  dynamical 
justification of  such a configuration for $J=3/2$ baryons is still missing, especially as a function of the radial quantum number $n$. 
Needless to say, that we do not have yet have experimental information on many $sss$ and any $ccc$ baryons.
 
 We start the paper by addressing the static three-quark potentials, and byevaluating these potentials  from the instanton liquid view of the
 vacuum fields. The results obtained will be compared to those calculated on the lattice, with rather good agreement as we will show.
However, for relevant distances these potentials  do not agree with the popular stringy $Y$ and $V$ models,
 but are closer to the so called "Ansatz A",  with half binary string  interactions.
 
This notwithstanding, we  still proceed  to the light front Hamiltonian, starting from a basic stringy  pictures as in the $Y$ and $V$ models.
  Following our analysis of  the mesons in~\cite{Shuryak:2021hng},  we derive and solve the light front Hamiltonian $H_{LF}$ 
  for baryons, by including the kinetic and 
  confining terms only. We will use Jacobi coordinates, exactly excluding the spurious CM motion, with the 
  ensuing quantum mechanics in six dimensions. 
  We will use the momentum representation, in which the kinetic and potential terms exchange
  roles.  The physical domain of the momenta will be certain triangles.
  As we will show, quantum mechanics on such manifolds is nontrivial but solvable.
  

 \section{Three-quark potentials}
 \subsection{Simplified models}
We start by enumerating the stringy models discussed in the literature.
 \\
\\
 {\bf 1. Y-model:}
 The simplest baryon configuration following from the quarkonium Cornell potential,
consists of a  perturbative Coulomb term plus a linear string potential. This is the
$Y$ (or ``star") model, with three strings going from quarks to 
 a certain point where the
``string junction" is located. 
\\
 {\bf 2. V-model:}
In the $V$-model,  one of the quark sits directly on the color junction,
and therefore there are only two strings in a baryon. 
 \\
 {\bf 3. A-model:}
This is  a somewhat  mysterious model,  which is nevertheless widely used by spectroscopists. 
We will call it A-model  as it corresponds to the following Ansatz
\be \label{eqn_A}
V_A={1 \over 2} \big( V_{q\bar q}(r_{12})+ V_{q\bar q}(r_{13}) +V_{q\bar q}(r_{23}) \big)
\ee
in which $V_{q\bar q}(r_{12})$ is the usual quarkonium potential, summed over all three pairs. Its mysterious element 
is the factor (1/2) in front, separating it from the so called $\Delta$ model. It
is justified at small distances, as the ratio of the perturbative color Casimir for $q \bar q$ and $qq$ pairs in baryons.   
At  large distances,   the potential is nonperturbative, and (\ref{eqn_A}) subsumes Casimir scaling.

Theoretical arguments 
using the  {\em vortex piercing} picture of confinement, have been  given in to support it~\cite{Cornwall:1996xr}, 
which carry also to large  $N_c$
 $$\frac 12 \rightarrow \frac 1{N_c-1}$$ much like the one-gluon exchange. Indeed, we know
on general ground that for $N_c=2$, mesons and diquarks should have the same potential. Also, for  large $N_c$,
the $qq$ force is down by $1/N_c$ in comparison to the $q\bar q$ force.  So perhaps this factor is 
in fact  correct for $all$ nonperturbative effects. 
\\
\\
These models provide certain predictions which vary depending on the geometry of the three quark locations.
We will discuss those as we proceed, and compare them with both numerically obtained 
potentials from lattice simulations, as well as with  our calculations using a  "dense instanton liquid" vacuum
model.
 
 \subsection{ Lattice  three-quark potentials  at large distances}
 If three quarks are $static$, their interaction can be evaluated using the correlators of three  Wilson
 lines $$\langle {\bf 1}-{\bf W}(\vec r_1) {\bf W}(\vec r_2) {\bf W}(\vec r_3)\rangle\sim e^{-V(\vec r_1,\vec r_2,\vec r_3)\tau}$$ running in the Euclidean time direction. In the case of 
 quarkonia  the color indices in $\langle {\bf W} {\bf W}^\dagger\rangle$
 can either be convoluted in a single trace, or a double
 trace (Polyakov loops). 
  The six color indices in ${\bf W}$ (not shown) can be convoluted in many more ways, and sandwiched with the initial and final
color wave functions of the baryon $\sim \epsilon^{abc}$. If the Euclidean time $\tau$ is infinitely long, they should all
lead to the same potential.

The three-quark static potentials have been assessed on the lattice  since the  1980$^\prime$s. For definiteness we will
use the  data compiled in  the Appendix in~\cite{Koma:2017hcm},
 for different geometries of the
quark locations. As these authors have shown, none of the models listed above provide a fully satisfactory fit to the data.
Nevertheless, they contain important lessons, few of which we would like to formulate in this section. 

The three static quarks can be arranged in triangles of different sizes and shapes.
Let us start with the most symmetric setting, the {\em equilateral} triangle. Following \cite{Koma:2017hcm}, 
three quarks are assigned to points $(x,0,0),(0,x,0),(0,0,x)$, and will be characterized by this distance $x$. All three sides
are $\sqrt{2}x$, but all plots below will be given as a  function of $x$. 

\begin{figure}[htbp]
\begin{center}
\includegraphics[width=6cm]{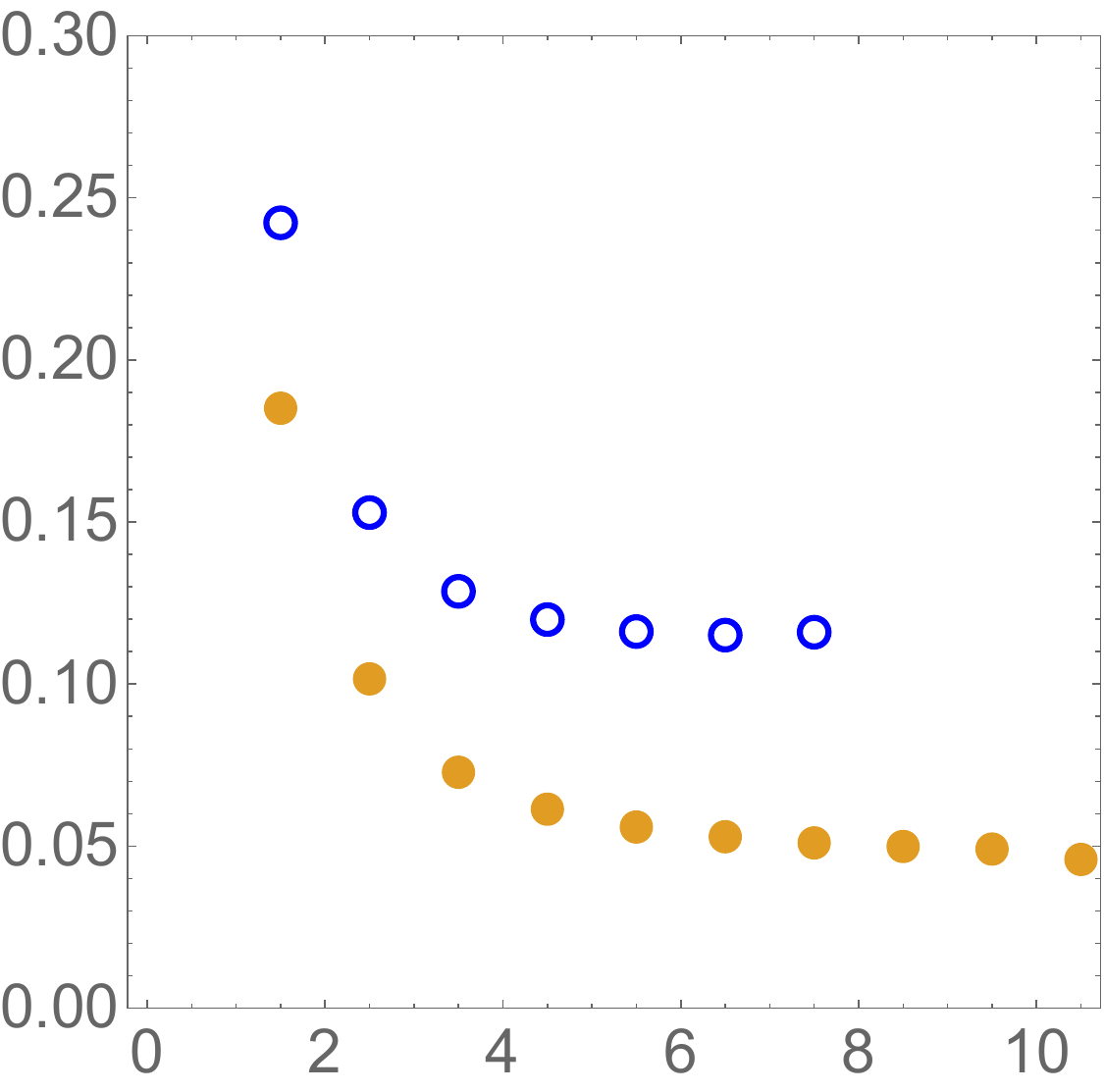}
\caption{Forces $dV/dx$ versus $x$ in lattice units, calculated from the lattice potentials as in~\cite{Koma:2017hcm}.
The open points are for an equilateral triangle, and the closed points are for a linear quark-diquark setting (the same as
the standard quark-antiquark one).
}
\label{fig_3qforces}
\end{center}
\end{figure}

Furthermore, let us focus on the large-$x$ limit, for which the differences between models 
is most obvious.  In order to exclude constant terms (possibly
depending on the triangle geometry),  we plot in Fig.\ref{fig_3qforces} the forces $dV/dx$ rather than the potentials.
They are compared to the quark-diquark potential (coinciding with the quarkonium potential as shown in the same paper). 
Indeed, at large distances the forces asymptote constants.

(This is expected for confining flux tubes. However, in the real world (or lattice simulations with dynamical quarks) 
we expect  ``avoided crossing phenomenon" to happen, whereby sufficiently long flux tubes break by 
production of a light quark-antiquark pair, so that  the growing potential turns into a constant at large distances.   
In fact, the real situation is more complex, as was understood long ago for crossing of molecular levels:
as shown by Landau, Zener and others in 1932,  the probabilities to follow or switch levels
depend on the velocity with which the crossing is approached. The more general issue of mixing between
baryons and pentaquarks is being discussed in spectroscopy. Using light front wave functions, one of us
\cite{Shuryak:2019zhv} evaluated this mixing in order to calculate PDFs of ``sea antiquarks". 
 In this paper we will ignore  the level crossing in the potentials.)

The constants in front of the linear terms depend on the geometry of the quark locations, and are 
different for different models. We will discuss three models and three geometries in the next subsection.
Here, we start with the well-known  $Y$-model, with three strings going from the quarks to the center of mass of the triangle, where the
``string junction" is located. In this model the force should be three times the string tension times the ratio of
the distance to the center in units of $x$. For  an equilateral triangle it  is
\be {dV_Y \over dx}= 3\sigma \sqrt{2\over 3}=\sqrt{6}\sigma
\ee
In the $V$-model, one of the quark sits directly on the color junction. 
In this case there are $two$ strings rather than three. For the particular triangle under consideration,
we have
 \be {dV_V \over dx}= 2\sigma \sqrt{2}=\sqrt{8}\sigma
\ee
Finally, for the $A$-model, we have the smallest force
\be {dV_A \over dx}= \frac 12 \,3\sigma \sqrt{2}
\ee
with $\sigma_A/\sigma_Y\approx 0.866<1$, whatever the triangular configuration.

In Fig.\ref{fig_3qforces}  the filled points correspond  to the string tension $\sigma$ for a quark-diquark (identical to quark-antiquark), using the
same lattice configurations. For the $Y$-mode, the ratio is
 \be {\sigma \over \sigma_Y} ={\sigma\over \sqrt{6}\sigma}\approx 0.854 \ee
 instead of 1.  For the  $V$-model we have  $\sqrt{8/6}\approx 1.15$, which is even larger.  In the $A$-model, it is comparable to $\sigma$
\bea
\frac \sigma{\sigma_A}\approx \frac{0.854}{0.866}\approx 1
\eea
 
The first lesson about the three-quark lattice potentials, is that the  string-based $Y,V$ models over-predict the forces at large distances, while
the  ansatz $A$-model  reproduces the force  (for an equilateral triangle).  Of course, we should consider other geometries, and probe
 intermediate distances as well, before ruling it as a succesful model fot the three-quark potential.
 
 The second lesson  is that some geometries are better for model separation than others. Indeed, another triangular limit
  in which one corner is far from the two others (including the quark-diquark picture), yields a single string tension times the longest
  distance for model-$Y$, which is the same as half the string tension times the two longest distances for model-$A$.

\subsection{Three-quark static potentials from the instanton vacuum}
Before proceeding to the static three-quark potentials from the three models presented earlier, using instantons, 
we  first put forth few remarks that suggest why model A would be favored.

It is a very important issue in the theory of few body quantum states (e.g. in nuclear physics tritium and ${\rm He}^3$) 
to separate effects  of well-constrained
two-body forces from ``truly three-body"  ones, which are then extracted from fits to experiment.
In  baryons made of three quarks,  we first observe 
that the color wave function $\epsilon^{abc}$ involves all three colors, while the instanton fields are
$SU(2)$ valued  in the space of three colors. This suggests that the main potential is best
approximated  by binary potentials. Among the models we presented, only Ansatz A (\ref{eqn_A})
has such a form.

Furthermore, since the instanton fields do not generate confinement, the pertinent potentials asymptote  constants, for the binary potential
$V(r_{ij}\rightarrow  \infty)  =2\Delta M$, and for the triple potential  $3 \Delta M$, etc. Since there are three
binary potentials, these large distance limits can only be reconciled if there is a factor $(1/2)$ in front, as
in Ansatz A.

The instanton-induced potentials between three static quarks  can be computed using the same
expressions for Wilson lines as was traditionally used for a  binary potential.
The  $SU(2)$ part of the Wilson line can be expressed using Pauli (rather than Gell-Mann) matrices $\tau^i$
\bea
{\bf W}^a_{lb}=\big(c_l\,{\bf 1}-i(\vec \tau\cdot \vec n_l)\,s_l\big)^a_b,\,\,\, a,b=1,2; l=1,2,3\nonumber\\
\eea
with  trigonometric functions involving  color rotation angles, that depend on the 3D distances $\vec \gamma_i,i=1,2,3$ between the 
location of Wilson lines $\vec r_i$, and  the instanton center $\vec z$
\bea
\label{LOC}
c_i&\equiv&{\rm cos}\bigg(\pi -\frac{\pi\gamma_i}{\sqrt{\gamma_i^2+\rho^2}}\bigg)\nonumber\\
s_i&\equiv &{\rm sin}\bigg(\pi -\frac{\pi\gamma_i}{\sqrt{\gamma_i^2+\rho^2}}\bigg)\nonumber\\
\gamma^2_i&=& 
(\vec r_i-\vec z)^2, \,\,\, \vec n_i=\vec \gamma_i/ |\vec \gamma_i |
\eea
We note that at large distances $|\gamma_i | \gg \rho$,  the phases vanish. At small distances they go to $\pi$,
with cosines  set to $-1$, with a fipped quark color direction. Since standard instanton does not act on a quark with the third color, this $2\times 2$
matrix should be  extended trivially (by 1)  to $3\times 3$.

 To properly model the color-isotropic  vacuum, 
  one should include some random $SU(3)$ matrices $U$. which rotate the  instanton fields from their standard $SU(2)$ plane,
  to an arbitrary  plane in $SU(3)$. These matrices should then be 
 averaged  using the Haar measure of the SU(3)  group (see Appendix  \ref{sec_Weingarten} for more details)
\bea
\label{3WP1}
\int dU (U^{a_1}_{i_1} {\bf W}^{i_1}_{1j_1} U^{\dagger j_1}_{b_1})(U^{a_2}_{i_2} {\bf W}^{i_2}_{2j_2} U^{\dagger j_2}_{b_2})(U^{a_3}_{i_3} {\bf W}^{i_3}_{1j_3} U^{\dagger j_3}_{b_3})\nonumber\\
\eea

 For the instanton-induced potential we thus get the following expression
\bea
\label{3WP3}
V&=&\frac{2 n_{I+\bar I}}{N_c}\int d^3z
\bigg[(1-c_1c_2c_3) \delta^{a_1}_{b_1}\delta^{a_2}_{b_2}\delta^{a_3}_{b_3}  \nonumber \\
&+&\frac{N_c^2}{(N_c^2-1)}\,c_1 s_2 s_3 n_2\cdot n_3 \delta^{a_1}_{b_1}\bigg(\frac 12 \lambda_2^B\bigg)^{a_2}_{b_2}\bigg(\frac 12 \lambda_3^B\bigg)^{a_3}_{b_3} \nonumber \\
&+&{\rm 2\, perm.}\bigg] 
\eea
where $n_{I+\bar I}$ is the 4D instanton plus antiinstanton  density. This is valid for any color states of the three quarks,. For a color singlet three-quark
baryon, one can either use its antisymmetric wave function  $\sim\epsilon^{abc}$, or  set the quark colors as  $a_1=b_1=1,a_2=b_2=2,a_3=b_3=3$ .

\begin{figure}[t]
\begin{center}
\includegraphics[width=5cm]{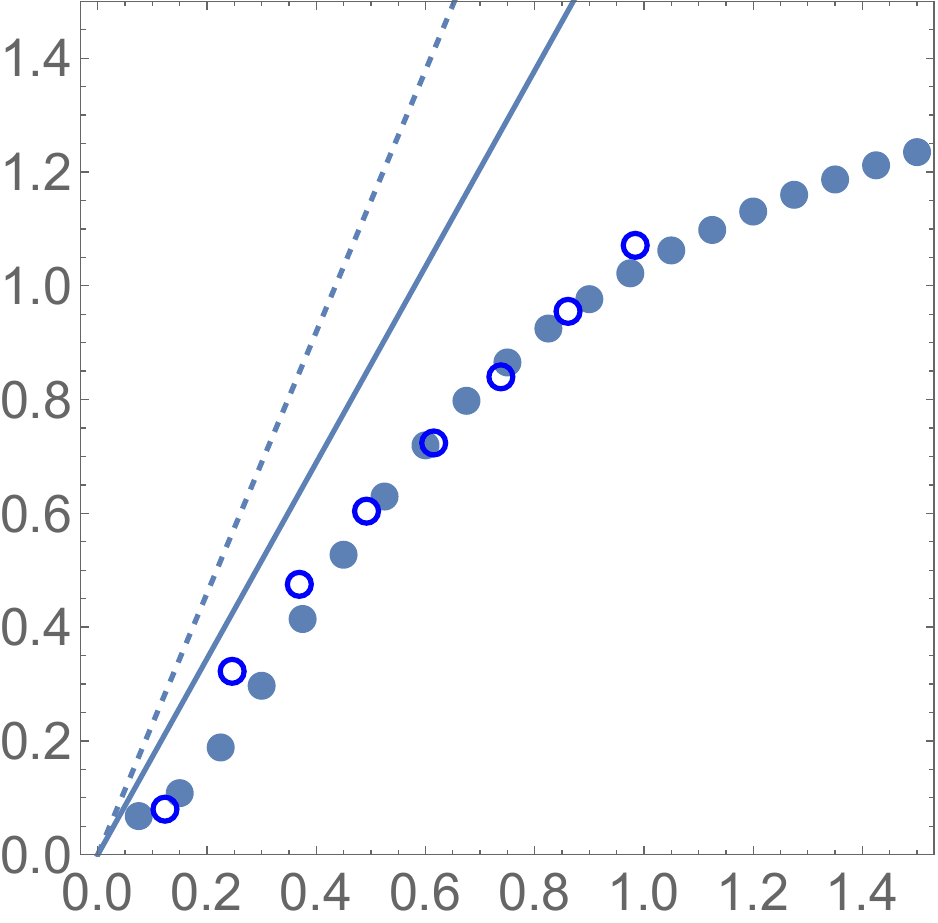}
\includegraphics[width=5cm]{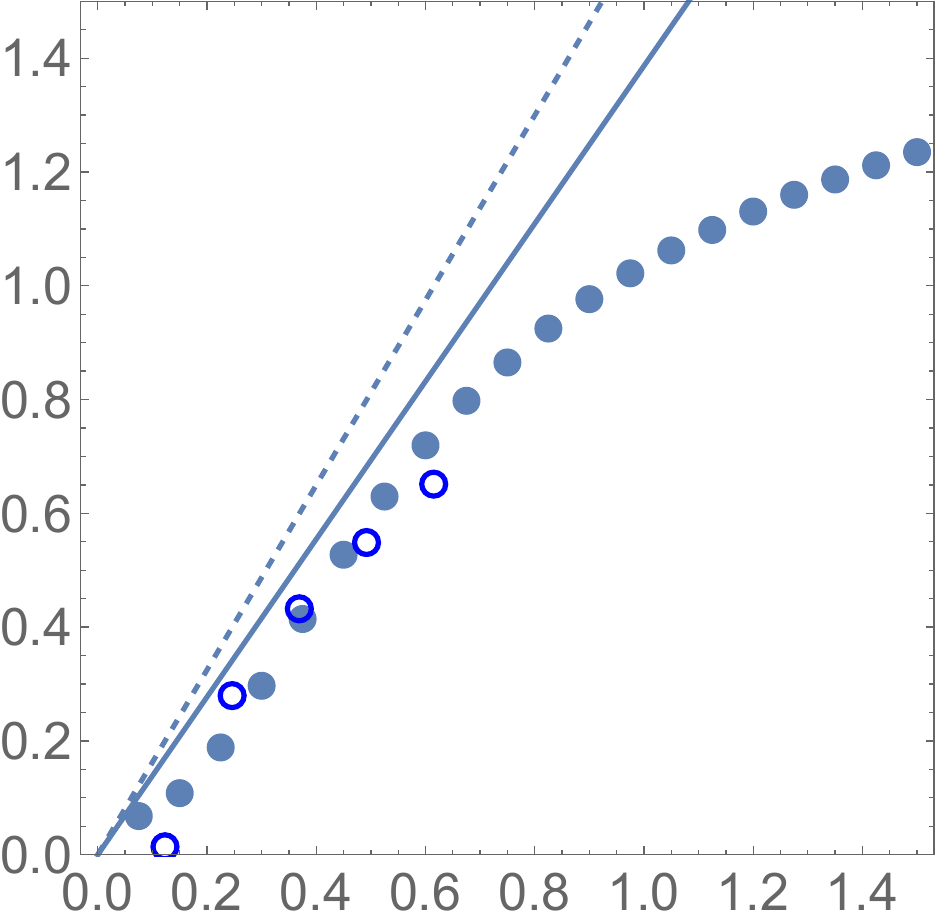}
\includegraphics[width=5cm]{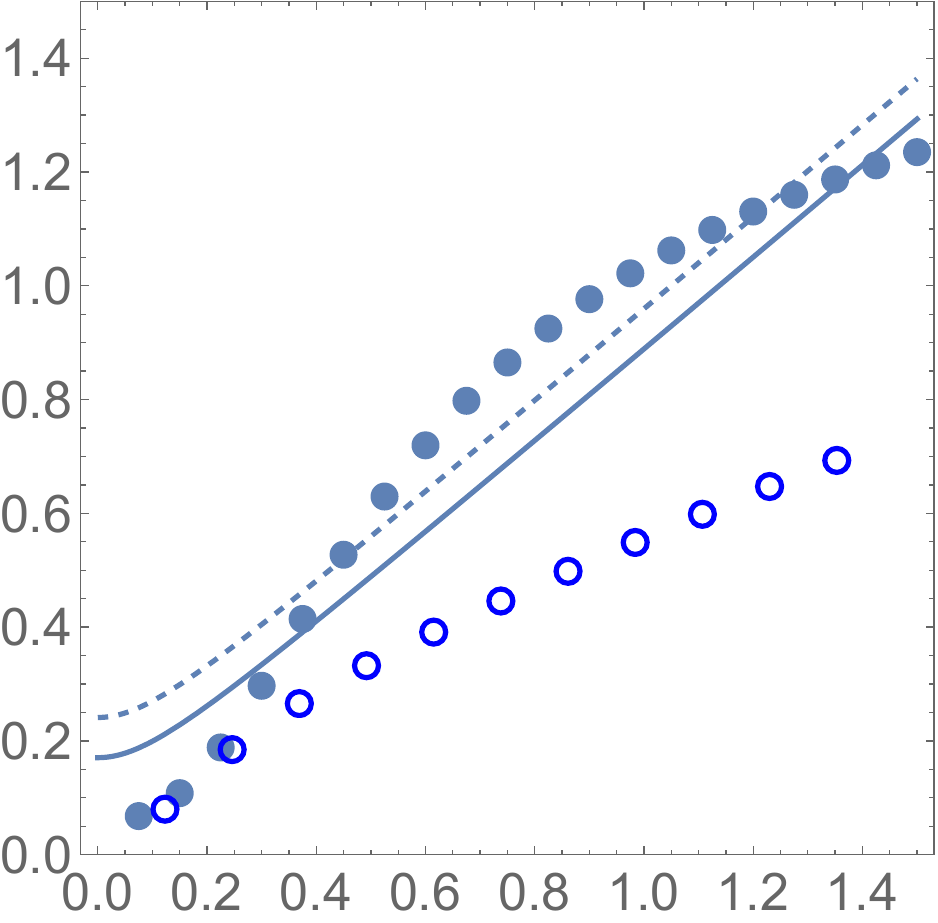}
\caption{The three-quark potentials $V\, (GeV)$ versus $x \, (fm)$  defined in (\ref{eqn_x}), for "equilateral", "direct"
and "long" triangles, top to bottom. The open points correspond to lattice simulation 
,  the closed points to our calculation in the ``dense instanton liquid" model. Linear predictions of the models $V$ and $A$ are shown
by the dashed and solid lines, respectively.
}
\label{fig_3q_pot}
\end{center}
\end{figure}

The results of the calculation of the instanton-induced potentials,  for the preceding three triangular shapes  are shown 
as filled circles in
Fig.\ref{fig_3q_pot} .  The instanton size was taken to be the usual $\rho=\frac 13$ fm,  and the 
density parameter $\kappa=\pi^2 n_{I+\bar I} \rho^4=1$ ($^{\prime\prime}$dense instanton liquid$^{\prime\prime}$). 
So the calculation has only one additive free parameter, which we have set  by requiring all potentials to  vanish for small triangles.
One can see that the shape of the potentials and their magnitude are rather independent of the shape of the triangles.
In all cases the potential at large x becomes close to $3\Delta M$, for  three independent quarks.

The potential is plotted  in Fig.\ref{fig_3q_pot}  versus $x$ defined via quark positions as follows (same  as on the lattice)
\ba  \label{eqn_x}
r_i&=&  (x,0,0) (0,x,0),(0,0,x)\,\,\,\,          "equilateral"\\
r_i&=&  (x,0,0) (0,x,0),(0,0,0)\,\,\,\,           "direct"\\
r_i&=&  (x,0,0) (0,a,0),(0,0,a)\,\,\,\,          "long" \ea
 In the plots,
 the dashed and solid straight lines refer to the  linear predictions of the models $V$ and $A$, discussed at the beginning of this section,  respectively. They correspond to 
\ba 
\label{XSLOPES}
V_V&=&\sqrt{8}\sigma x,\,\,\, V_A={3 \over \sqrt{2}}\sigma x \,\,\,  "equilateral" \nonumber \\
V_V&=&2\sigma x,\,\,\, V_A=\sigma x(1+1/\sqrt{2}) \,\,\,   "direct"  \nonumber\\
V_V&=&\sigma (\sqrt{x^2+a^2}+a\sqrt{2}), \\
 V_A&=&\sigma ( \sqrt{x^2+a^2}+a/\sqrt{2})\,\,\,  "long"  \nonumber 
 \ea

Since the constants in the potentials are not well defined, (\ref{XSLOPES}) are shown for comparison to the "force" (the slopes),
with the  instanton-induced (filled circles) and lattice potentials (open circles).
 Note that here we used the  ``empirical" string tension $\sigma=(0.4\, GeV)^2$,   deduced from the Regge slope $\alpha^\prime$. The 
corresponding lattice value is smaller by as much as 30\% (perhaps due to the limited size of the lattice, or other uncertainties in the scale). Within this accuracy range,
 we conclude that the slope predicted by model A is in {\em crude agreement} with the  instanton-induced potentials.
 
  The open circles in these plots are  from the lattice simulation Tables in  \cite{Koma:2017hcm}, the lattice site size
is $a=0.123\, fm$. Since a pointlike charge energy require renormalization, we also shifted them downward by a constant.
One can see that the instanton-induced and lattice potentials agree quite well for two former triangles, but for the ``long"
one even the slope does not agree.
 Note  however that: (i) the smallest side of the triangle in this case is not 
 growing with $x$ but remains constant and small  $\sqrt{2}a=0.174\, fm$; (ii) that this potentials goes roughly to $2\Delta M$
 rather than  $3\Delta M$ as in the other cases. The lattice seems not to resolve the quark pair at this distance,
 while our continuous instanton formulae do. To resolve the issue one perhaps need an instanton model with a
 realistic size distribution, and also finer lattices. 

{\bf In conclusion}: Good agreement is shown between the instanton-induced and
lattice potentials for large-enough triangles, but not for ``long" ones.  Of all the three models Y,V,A discussed, 
model-A t seems to be closer to the evaluated instanton-induced potentials, even for ``long" geometry.

 \section{Preliminaries}
 \subsection{Jacobi coordinates}
The LFWFs for baryons discussed in literature so far consider quarks as independent, so those are a function of
9 coordinates (or 9 momenta).  The spurious CM motion is implicitly present: in some cases the corresponding
energy is subtracted, but corrections to the wave functions are simply ignored.
 
 However,  there is no need for this.  Exact kinematics with appropriate conditions can be satisfied by a well known
 change of variables, widely used in many few-body applications. Total momentum is subject to three conditions
 \bea 
&& \vec p_\perp^{tot}=\vec p_{1\perp}+\vec p_{2\perp}+\vec p_{3\perp}=0\nonumber\\
&& x_1+x_2+x_3=1 
 \ea 
so the system is in fact 6-dimensional. 

The main idea of the approach we use is to work in momentum representation, with kinetic term
of the Hamiltonian treated as a ``potential", and the confining part (in which coordinates are used
as derivative over momenta $\vec r_i=i{\partial / \partial \vec p_i}$

For transverse momenta we introduce two  (slightly modified) Jacobi momenta variables
 \be \label{eqn_Jacobi}
\vec p_{\rho\perp}={1 \over \sqrt{2}} (\vec p_{1\perp}-\vec p_{2\perp}), \, \, \vec p_{\lambda\perp} ={1 \over \sqrt{6}}(\vec p_{1\perp}+\vec p_{2\perp}-2 \vec p_{3\perp}) \ee
 in term of which
  \ba \label{eqn_inverse}
\vec p_{1\perp} &=&  ( \sqrt{6}\vec p_{\lambda\perp}+ 3 \sqrt{2} \vec p_{\rho\perp})/6, \nonumber \\
\vec  p_{2\perp} &=& ( \sqrt{6} \vec p_{\lambda\perp}- 3 \sqrt{2} \vec p_{\rho\perp})/6, \nonumber \\
 \vec p_{3\perp} &=& - \sqrt{6}  \vec p_{\lambda\perp}/3
 \ea
Now the total transverse momentum $\vec p_{tot}= \vec p_1+\vec p_2+\vec p_3$ vanishes automatically.
 
 The longitudinal momentum fractions are defined similarly
 \ba \label{eqn_Jacobi_x}
x_1&=&(\sqrt{6}\lambda+3\sqrt{2}\rho+2X)/6 \nonumber \\
x_2&=&(\sqrt{6}\lambda-3\sqrt{2}\rho+2X)/6 \nonumber \\
x_3&=&(-\sqrt{6}\lambda+X)/3 \ea
Note that $X=x_1+x_2+x_3$:  unlike in the transverse direction for which $X=0$, here $X$ should be set to 1. 
Therefore, the physical domain is the three-dimensional cube $x_i\in[0,1]$  {\em cut by a plane} $X=1$, leaving as a
physical domain a triangle in $\lambda,
\rho$ coordinates  inside which  the parton fractions are all positive $x_i>0$. The three corners of the triangles
 correspond to parton configurations with
  one quark with its fraction  being 1,  and the  two others zero. The longitudinal part of  LFWF should therefore be defined  
{\em on this triangle}.

After the wave functions in momentum representation are defined, one can reconstruct their versions in coordinate 
 representations by the usual Fourier transform. The coordinates conjugated to $\vec p_\rho,\vec p_\lambda$, will be
 referred to  as $\vec r_\rho,\vec r_\lambda$. 
 
 The confing part of the Hamiltonian in its simples form can be re-written using the einbein trick used 
 in~\cite{Shuryak:2021hng,Shuryak:2021mlh} and reminded in Appendix \ref{sec_einbine}.  For the  $Y$ (or star) model this amounts to 
 \begin{widetext}
 \ba 
 \label{EEX1}
 \sigma_T(| \vec r_1 |+| \vec r_2 |+| \vec r_3 |) \rightarrow
 \frac{ \sigma_T}2 \bigg( \frac 3a + a  (\vec r_1^2+\vec r_2^2+\vec r_3^2) \bigg)
 \rightarrow  \frac{ \sigma_T}2 \bigg( \frac 3a + a  (\vec r_\lambda^2+\vec r_\rho^2) \bigg)
 \ea
 with all einbein parameters set equal to $a$ by saddle point. For Ansatz A (\ref{eqn_A}) this amounts to
  \ba 
  \label{EEX2}
  {\sigma_T \over 2} (r_{12} +r_{23} +  r_{13} ) \rightarrow \frac{\sigma_T}4  \bigg(\frac 3a +a  ( r_{12}^2+ r_{23}^2+ r_{13}^2) \bigg)
 \rightarrow   \frac{ \sigma_T}4 \bigg( \frac 3a + 3a  (\vec r_\lambda^2+\vec r_\rho^2) \bigg)
 \ea
 \end{widetext}
 Note that when $a$ is to be eliminated by minimization, the minimum is at $a^*=\sqrt{3/(r_\lambda^2+r_\rho^2)}$
 in  (\ref{EEX1}), but $a^*=\sqrt{1/(r_\lambda^2+r_\rho^2)}$ in (\ref{EEX2}).
 Substituting those values back to original expressions, one finds that
  confining potentials in (\ref{EEX1}) in
 Jacobi coordinates  is the same as  (\ref{EEX2}) 
 after the following rescaling of the string tension
 \bea
 \label{RESCALING}
 \sigma_T\rightarrow \frac{\sqrt 3}2 \sigma_T\,.
 \eea
Note that rescaling factor  $ \frac{\sqrt 3}2 \approx 0.866<1$, so the tension (and thus all baryonic masses)
corresponding to model A are lower than in model Y, as expected.

The quantum replacement  $$\vec r_\lambda \rightarrow i {\partial \over \partial \vec p_\lambda},\,\,\,\vec r_\rho\rightarrow i {\partial \over \partial \vec p_\rho} $$
   yields  a 6-D Laplacian  in $\vec p_\lambda,\vec p_\rho$, for both the Y and A models.  
  The kinetic term contains sum of corresponding momenta squared: so, in this approximation,
  there appears 6-D spherical symmetry in Jacobi coordinates.

 \subsection{Relativisitic semiclassical quantization in the rest frame} \label{sec_semiclassics}
Before we move to the light front, however, we present
 our preliminary study of the problem in the rest frame. We  will present in details the
  $Y$ or ``star configuration",  with three strings connected  to a junction.
  The junction will be assumed static and located in the CM for equal masses.
 The results for model A will be briefly quoted.  In the rest frame the baryon with zero orbital momentum  is spherically symmetric.

Heavy quarks can be treated via the Schroedinger equation, but for  light quarks  
their effective masses and momenta are comparable, so nonrelativistic approximation is invalid. 
need to be addressed  differently.
This distinction however can be avoided in the semi-classical approach we will use here (and of course
the light front approach is the same for light and heavy quarks).
The  Hamiltonian  for the $Y$ configuration is 


\bea
\label{HRF}
H=\frac 1{2m}\big(\vec p_\lambda^2+\vec p_{\rho}^2\big)+\sigma_T \sum_{i=1}^3|\vec{r}_i|
+\bigg(\frac 32 \frac{m^2_Q}m+\frac 32 m\bigg)\nonumber\\
\eea
Here $m_Q$ is the quark mass, $\sigma_T$ the string tension, and $m=1/2e$ is the variational effective mass,
arising from the  einbein trick  used to unwind the relativistic square-root.  A similar Hamiltonian was obtained in~\cite{Simonov:1989ff},
using a different world-sheet embeding than the one we will present below (see (\ref{EMBEDDING})).
For the confining part, (\ref{EEX1}) gives
\bea
\label{REDUCTION}
\sigma_T\sum_{i=1}^3|\vec{r}_i|
\rightarrow \sqrt{3}\sigma_T(\vec r_\lambda^2+\vec r_\rho^2)^{\frac 12}
\eea
 (\ref{HRF}) simplifies to that of $one$ particle
in a $D=6$-dimensional space
\bea
\label{PMU}
H\rightarrow &&\,\,\, \frac 1{2m}\big(\vec p_\lambda^2+\vec p_{\rho}^2\big)\nonumber\\
&&+\sqrt{3}\sigma_T (\vec r_\lambda^2+\vec r_\rho^2)^{\frac 12}
+\bigg(\frac 32 \frac{m^2_Q}m+\frac 32 m\bigg)\nonumber\\
\rightarrow&&\,\,\, \frac {p_\mu^2}{2}+{\tilde{\sigma}_T}|Z_\mu| +\bigg(\frac 32 \frac{m^2_Q}m+\frac 32 m\bigg)
\eea
 (\ref{PMU}) describes a non-relativistic and linearly confined particle of variational mass $m$,
with coordinates $Z_\mu=(\lambda^i, \rho^i)$ in $D=6$ dimensions as per the last relation. We 
have rescaled 
the  coordinate $\sqrt{m}Z\rightarrow Z$ and  string tension  $\tilde{\sigma}_T=\sqrt{3}\sigma_T/\sqrt{m}$,
for convenience.



An estimate of the mass spectrum can be obtained using
the WKB approximation,

\bea
\int_{r_S}^{r_L}dr \bigg(2E-2\tilde\sigma_T r-\frac{l(l+D-2)}{r^2}\bigg)^{\frac 12}= \bigg(n+\frac 12\bigg) \pi\nonumber\\
\eea
with the end points $r_{L,S}$ solution to the cubic equation
$$2\tilde\sigma_Tr^3-2Er^2+l(l+D-2)=0$$
For zero orbital motion $l=0$, the WKB radial energy levels 
can be found to be
\bea
E_{n0}(m)=\bigg(\frac{3\pi}{2\sqrt{2}}\bigg)^{\frac 23}\bigg(n+\frac 12\bigg)^{\frac 23}\tilde\sigma_T^{\frac 23}
\equiv \frac{\tilde{E}_{0n}}{m^{\frac 13}}\nonumber\\
\eea
 Once combined with the extra terms in (\ref{PMU}), we  can carry the minimization in $m$,  and 
set its value at the minimum. The ensuing WKB radial mass spectrum of the star baryon $M_{n0}$ in the rest frame,
Reggeizes  for large $n$ $linearly$

\bea
\label{RNN}
\alpha^\prime M_{n0}^2\approx 2\sqrt{3}\, n
\label{REG1}
\eea
with $\alpha^\prime=1/2\pi\sigma_T$. 
We recall  that the meson  Regge trajectory  is $\alpha' M^2=1$. So, in the same units our
``star-shaped" baryons have a  slope  $2\sqrt{3}\approx 3.46$, compared to the mesons.
It is close but not equal to the number $3$, naively corresponding to the number of strings.
This  WKB radial Regge trajectory calculated in the rest frame, has similar but not identical slope to that 
derived from the light front (see  (\ref{M2NN})  below).


For large orbital excitations $l$, the motion is classical, and 
an estimate can be obtained by noting that for the confining potential 
the virial theorem gives 
$$E_{0l}\approx K+V=3K =\frac{3l^2}{2R^2}$$
with $R=(l^2/\tilde\sigma_T)^{\frac 13}$ fixed by the force equation.
After fixing $m$ by minimization,
the mass spectrum of the star baryon is seen to Reggeize linearly in large orbital momentum  $l$ as well, 
\bea
\label{RLL}
\alpha^\prime M_{0l}^2 \approx \frac{6}\pi \,l
\label{REG2}
\eea
 with a slope $6/\pi\approx 1.91$, so  the
linear Reggeization is not the same in $n$ and $l$! This is in disagreement with the experimental data for light baryons, as we have
demonstrated above for the isobars.

The results for model $A$ follows from those for model $Y$ through the rescaling (\ref{RESCALING}). In particular, the Reggeized 
trajectories in the semi-classical approximation are 
\bea
\alpha^\prime M_{n0}^2&\approx& 3\, n\nonumber\\
\alpha^\prime M_{0l}^2 &\approx& \frac{3\sqrt 3}\pi \,l
\eea
in comparison to (\ref{RNN}) and (\ref{RLL}), respectively.

 \section{The Hamiltonian on the light-front }  
 

The kinetic part of the LF Hamiltonian has the form 
$$\sum_i {\vec p_{i\perp}^2+m_i^2 \over 2p_{i\, long}}$$
in which the transverse and longitudinal momenta appear differently.
As in our previous papers, we rewrite  it in the following form
\bea 
&&{p_{1\perp}^2+m_Q^2 \over x_1}+{p_{2\perp}^2+m_Q^2 \over x_2}+{p_{3\perp}^2 +m_Q^2\over x_3}=  \nonumber \\
&& 3 \sum_i (p_{i\perp}^2+m_i^2 )  +\sum_i (p_{i\perp}^2+m_i^2)\bigg({1  \over x_i}-3\bigg) \nonumber
\eea
The first quadratic term in the last line lead to a {\em transverse oscillator}, 
and the
 second term is called a {\em nonfactorizable potential} $\tilde V$,  it mixes transverse and longitudinal variables. It
will be included by different methods to be defined below.

The confining part of the LF Hamiltonian is built from terms linear in coordinates. For example 
$Y$ model with three strings going to the junction at the center has
  a sum of linear terms $$V_Y=\sigma_T \sum_{i=1}^3 | \vec r_i |$$
Like we did for mesons in our previous papers,   
 we again introduce {\em en einbine trick} with a variational parameter $a$ which  allows to
 re-write  the  linear potential as quadratic one. 

Furthermore, as in our previous papers, we use the Hamiltonian in the momentum representation. Therefore
the coordinate vectors are interpreted as $\vec r=i \partial / \partial\vec  p$, and
therefore  the
confining part $\sim vec r^2$ will play the role normally attributed to the kinetic energy. Quadratic confinement
thus leads to a second order Schroedinger-like equation for the eigenfunctions. 

The same logics is applied to the transverse and longitudinal coordinates $\vec r_\perp, r_{long}$,
so the immediate task is to write the Laplacian operator,  both in Jacobi coordinates in
transverse and in our curved map (\ref{eqn_long_map}). Both tasks are performed,
as explained in Appendix~\ref{sec_app_st}.

Let us focus for now on longitudinal momenta. In variable $\lambda,\rho,X$ (\ref{eqn_Jacobi_x})
the line element defining the metric
tensor in the new coordinates, is diagonal and simple
\be dl^2=d\lambda^2+d\rho^2+dX^2/3 \ee
The  Laplacian (which we encounter in the confining term of the Hamiltonian) in the original coordinates also takes a simple form
\be \nabla^2= \sum_i {\partial^2 \over \partial x_i^2} \rightarrow {\partial^2 \over \partial \lambda^2}+
{\partial^2 \over \partial \rho^2}+3{\partial^2 \over \partial X^2} \ee
Since we work on constant $X=1$ we need only the first two terms. 

   Therefore, the first problem we encounter is to define eigenfunctions of the Laplacian
    on the triangular physical domain in the $\lambda-\rho$ variable.
 As we will show below, for the equilateral triangle this problem can in fact be solved analytically.  

\begin{figure}[h]
\begin{center}
\includegraphics[width=6cm]{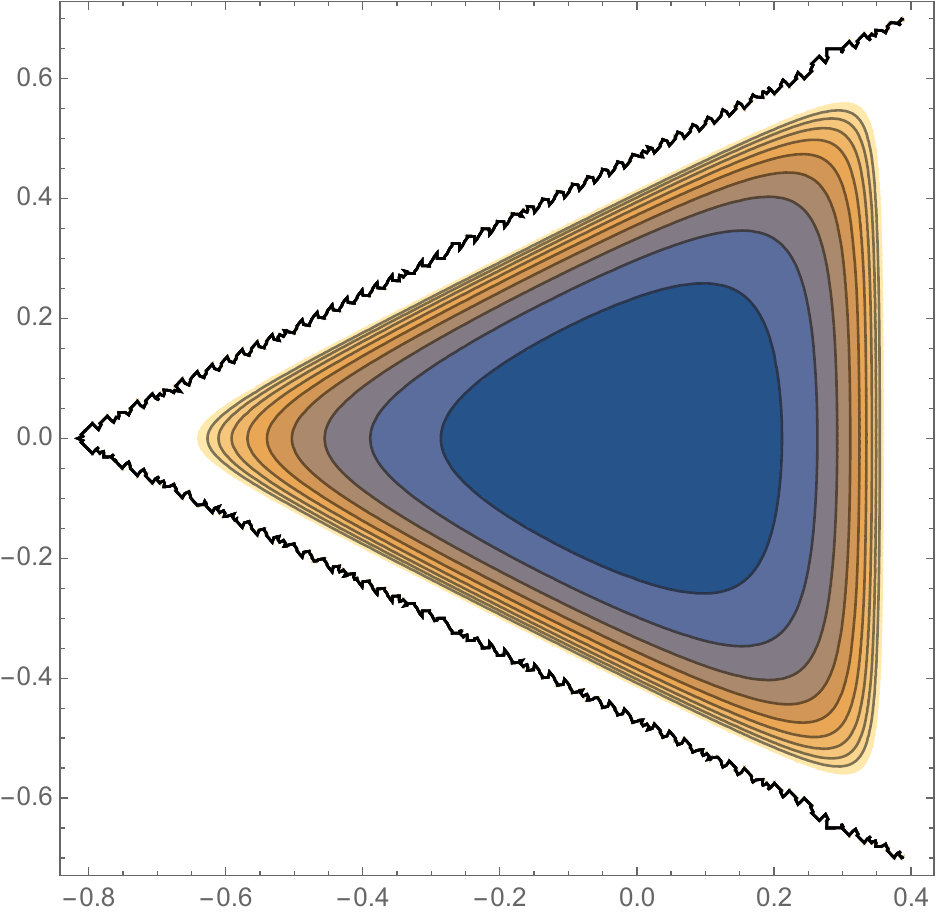}
\includegraphics[width=1cm]{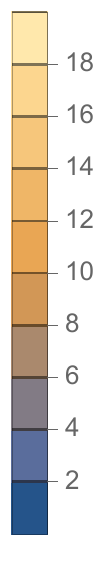}
\caption{The contour plot of the ``triangular cup" potential $V(\lambda,\rho)$ on $\lambda,\rho$ plot.}
\label{fig_V_in_Jacobi}
\end{center}
\end{figure}

The  main difficulty is related with the non-factorizable potential $\tilde V$. Its structure is schematically
 given by the  combination \be \tilde V \sim \bigg(\frac 1{x_1}+\frac 1{x_2}+\frac 1{x_3} -9\bigg)\,, \ee 
 assuming ($(\vec p_i^\perp)^2+m_i^2)$ can be approximated by its average and factor out. The main feature of $\tilde V$
is that it is small near the center of the triangle, but becomes large
  at all boundaries, see e.g. its contour plot 
 shown in Fig.\ref{fig_V_in_Jacobi}. Therefore we  call it a ``triangular cup".  The singular nature of $\tilde V$ at the boundaries, leads to divergences in matrix elements, unless the wave functions 
 vanish there. Therefore, the problem we set to solve must have Dirichlet boundary conditions $\psi_i(\lambda,\rho)=0$
 at the boundaries for all functions.
 
 Quantum mechanics on a triangle
 with the potential $\tilde V$ will be solved below, by two numerical methods.
 But before we do so, it is always useful to start with less accurate but much simpler variational method.

\begin{figure}[h!]
\begin{center}
\includegraphics[width=6cm]{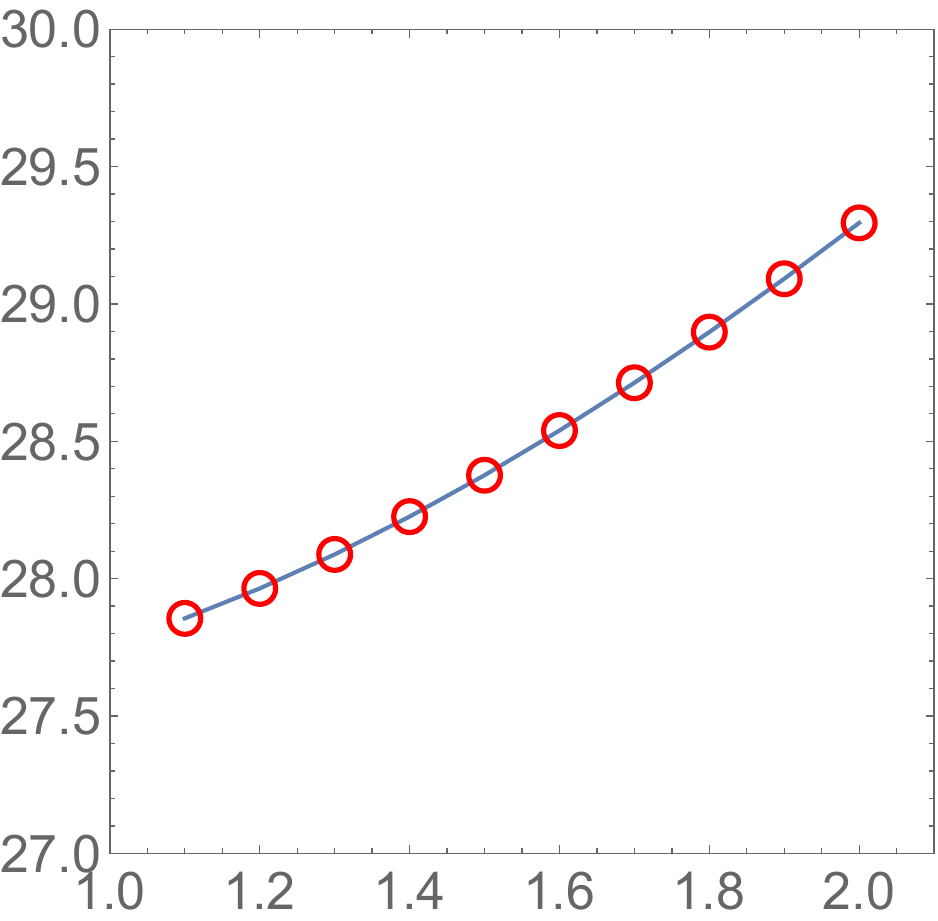}
\includegraphics[width=6cm]{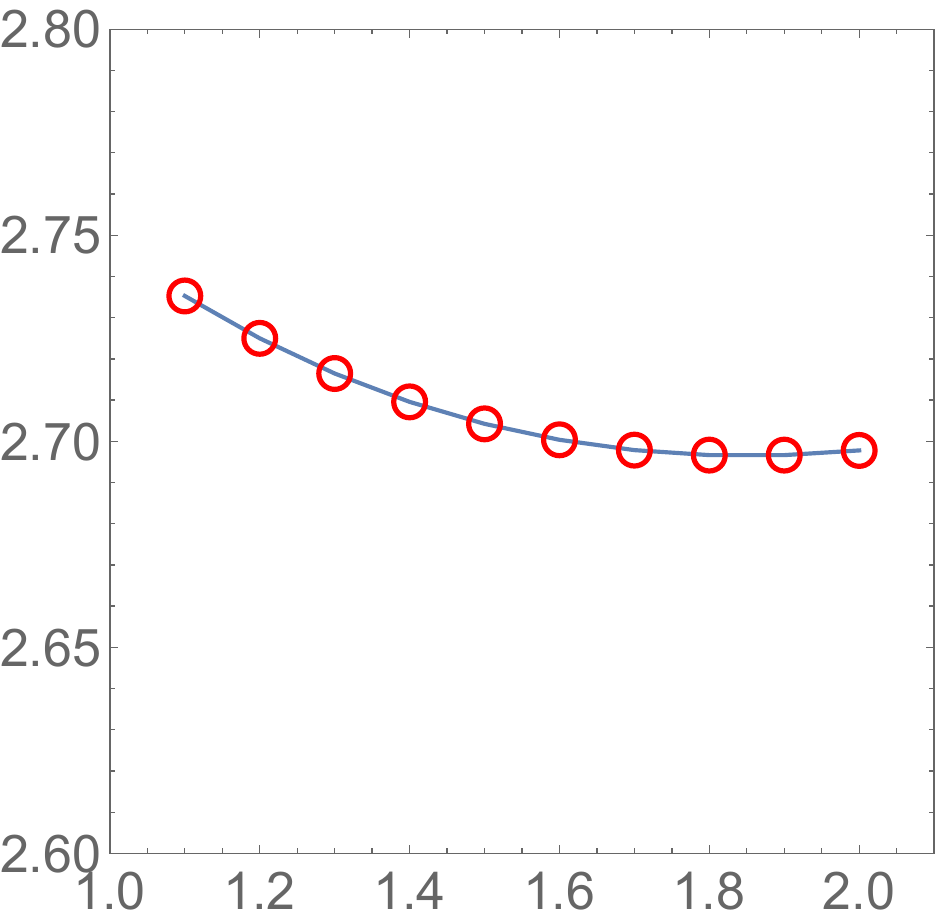}
\caption{The average Laplacian (upper) and $V(\lambda,\rho)$ (lower plot) versus the variational paremeter $A$. See text.}
\label{fig_L_V_av}
\end{center}
\end{figure}

To exclude divergences on the boundaries, the wave function 
should vanish, so we simply include  linear suppression factors and assume that
\be \label{JASTROW}\Psi( \lambda,\rho)=\big[ \prod_i x_i(\lambda,\rho)\big] \Phi( \lambda,\rho) \ee
with some regular $\Phi$ (This procedure is known in nuclear and condensed many-body physics, through the use
of Jastrow type wave functions). Let us then take this regular function to be  
a Gaussian  centered in the  triangle
\be \label{GAUSS}\Phi(\lambda,\rho)= exp\bigg( -A\bigg(\lambda^2 +\bigg(\rho - {1\over \sqrt{6}}\bigg)^2\bigg)\bigg) \ee
with a variational parameter $A$. We use  (\ref{JASTROW}),   evaluate  the average of the Laplacian 
and of the potential $V$, and plot the result 
as a function of $A$ in Fig~\ref{fig_L_V_av}.  As expected, increasing $A$ -- that is making the wave function
better localized near the center -- leads to a growth of the mean Laplacian and a decrease of the mean $V$.
Taking those two averages with proper
coefficients, one finds a minimum of the total Hamiltonian. 

\subsection{Longitudinal momentum fractions in factorizable coordinates}
The longitudinal motion can be  treated in a different way, by a nonlinear but factorizable maps
into a new set of variables. This mapping 
 was developed in~\cite{Shuryak:2019zhv} for any number of constituents,
in particular it was used for the 3 and 5 quark sectors of the baryons.

Let us parameterize the  three momentum fractions of the quarks,   using the following three parameters $s,t,u$
\ba \label{eqn_long_map}
x_1&=&u\big({1+s \over 2}\big)\big( {1+t \over 2}\big) , \nonumber \\
 x_2&=&u\big( {1-s \over 2} \big)\big({1+t \over 2} \big), \nonumber \\
x_3&=& u\big({1-t \over 2} \big), \ea 
The longitudinal momentum 
constraint  $x_1+x_2+x_3=u$, will be enforced later $u\rightarrow 1$.

\begin{figure}[htbp]
\begin{center}
\includegraphics[width=6cm]{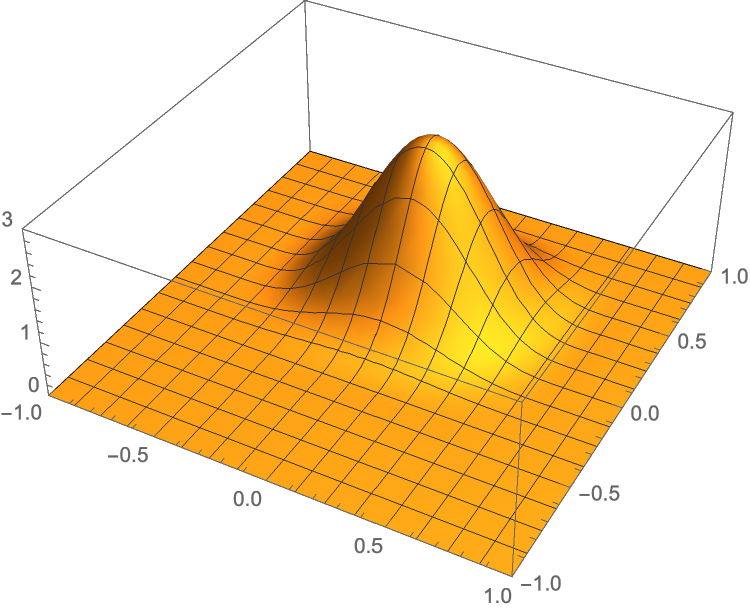}
\caption{The wave function of $\Delta(3/2)$ baryon, in $s,t$ coordinates, from \cite{Shuryak:2019zhv}. } 
\label{fig_Delta_s_t}
\end{center}
\end{figure}

The  inverse map, explains better the meaning of  $s,t$ as ``asymmetries"
\ba s&=&{x_1-x_2 \over x_1+x_2}, \nonumber \\
t&=&{x_1+x_2-x_3 \over x_1+x_2+x_3}, \nonumber \\
u&=& x_1+x_2+x_3 \ea
The corresponding metric and Laplacian in  the  $s,t,u$ coordinates are listed  in Appendix~\ref{sec_basis}. 
The main point is that the physical domain of the $s,t$ variables is a $square$,  since both vary
between -1 and 1. With the help of appropriate Jacobi polynomials, one can have
factorized orthonormal basis functions, in terms of which the Hamiltonian matrix elements can be computed. 
With some model Hamiltonian (different from the one used in the present paper),
the mass and wave function for the lowest $\Delta$ states have been evaluated  in~\cite{Shuryak:2019zhv}, see
Fig.~\ref{fig_Delta_s_t}. We show it in order to compare with the wave functions to be derived below.
Note that the wave function is approximately Gaussian, with strong suppression
near the edges of the physical domain. (It  is rather different from that of the nucleon, see the original paper.)

In this paper we will not use the $s,t$ coordinates and the Jacobi polynomial basis. Yet we note,
that whatever coordinates or basis is used, one cannot simply invent a convenient
Hamiltonian in those coodinates, plus whatever motivations. In particular,  the Laplacian in the
original coordinates should be re-written using the pertinent
 expressions  from differential geometry. For the $s,t$ map given above, the Laplacian is involved and listed 
in Appendix~\ref{sec_app_st}.

\section{Nambu-Goto string and confinement}
\subsection{Confining light front Hamiltonian}


 Ignoring Coulomb and spin effects, we start by focusing on confinement by a relativistic string.
 The action in the first quantized form can be written as

\bea
\label{ACTION1}
S[\theta]=&&\int_0^T d\tau  \sum_{i=1}^3\bigg(e_i m_i^2+\frac 1{4e_i} \dot{x}_i^2\bigg)\\
&&+
\sigma_T\sum_{i=1}^3\int_0^T d\tau  \int_0^1d\sigma_i 
\sqrt{\dot{X}^2_i{X}_i^{\prime 2}-(\dot{X}_i\cdot X_i^\prime)^2}\nonumber 
\eea
In the first term, describing endpoint masses, we use the  ``einbein trick" which we will use
consistently throughout these papers to get rid of unwanted square roots. Note that if
one performs  minimization with respect to the three einbein parameters $e_i$,  it yields back the standard free relativistic
action for massive particles (in Euclidean signature). 

The string world-sheet action in the Nambu-Goto action includes derivatives over internal coordinates $\tau,\sigma$
shown by a dot and prime, respectively. The world-sheets themselves
can be described by  the
so called ``ruled surfaces", parametrized by
\bea
\label{EMBEDDING}
X_i^\mu(\tau, \sigma_i; \theta)&=&z^\mu(\tau, \theta)+\sigma_i r_i^\mu \nonumber \\
 r_i^\mu &=& (r_{i\perp}, r_{i3}, 0) \\
z^\mu(\tau, \theta)&=&(0_\perp, {\rm sin}\theta\tau, {\rm cos}\theta \tau)  \nonumber
\eea
and $z^\mu(\tau, \theta)$  being the world-line of the string junction. (Our notations for the 
coordinates  are  1,2 for transverse, 3 for  longitudinal beam direction , and 4 for  time.)

For baryons in the so-called {\it star configuration}, the string junction and the end-points follow parallel
trajectories,  sloped at angle $\theta$ with respect to the 4-direction. For $\theta=0$, the analysis  corresponds
to a star baryon in the  rest frame. For arbitrary $\theta$ with subsequent analytical continuation $\theta\rightarrow -i\chi$, the analysis
corresponds to a star baryon on the light front.

As already explained above, to factor out spurious motion of the center of mass, we use Jacobi coordinates. For equal  quark masses  
$$m_1=m_2=m_3=m_Q\ ,$$
the center of mass coincides with the location of the string junction $z^\mu$. Also, although the einbeins
are arbitrary and fixed only by minimization for the free part, symmetry suggests that the minima are equal or
$e_1=e_2=e_3=e$, with only $e$ to minimize, by steepest descent. This will be assumed throughout.

The specific form of the Jacobi coordinates for the end-points is
\bea
r^\mu_1&=&\frac 1{\sqrt{6}}r_\lambda^\mu+\frac 1{\sqrt{2}}r_\rho^\mu\nonumber\\
r^\mu_2&=&\frac 1{\sqrt{6}}r_\lambda^\mu-\frac 1{\sqrt{2}}r_\rho^\mu\nonumber\\
r^\mu_3&=& -\frac{\sqrt{2}}{\sqrt{3}}r_\lambda^\mu
\eea
with a kinetic contribution

\bea
\label{END1}
\int_0^Td\tau \bigg(3em_Q^2+\frac 3{4e}+\frac 1{4e}(\dot{r_\lambda}^2+\dot{r_\rho}^2)\bigg)
\eea
in  (\ref{ACTION1}). The Nambu-Goto string contribution is

\bea
\label{NB2}
\int_0^Td\tau\, {\sigma_T} \sum_{i=1}^3 |\xi_i(\theta)|
\eea
with the invariant distances
$$|\xi_i(\theta)|=(r_{i\perp}^2+{\rm cos}^2\theta r_{i3}^{2})^{\frac 12}\ ,$$ 
or, in the Jacobi coordinates
\bea
\label{NB3}
\xi^2_1(\theta)=&&\frac 16 r_{\lambda \perp}^2+\frac 12 r_{\rho\perp}^2\nonumber\\
&&+\frac 16 r_{\lambda\perp}\cdot r_{\rho \perp}
+{\rm cos}^2\theta\bigg(\frac 1{\sqrt{6}}r_{\lambda 3}+\frac 1{\sqrt{2}}r_{\rho 3}\bigg)^2\nonumber\\
\xi^2_2(\theta)=&&\frac 16 r_{\lambda \perp}^2+\frac 12 r_{\rho \perp}^2\nonumber\\
&&-\frac 16r_{\lambda \perp}\cdot r_{\rho \perp}
+{\rm cos}^2\theta\bigg(\frac 1{\sqrt{6}} r_{\lambda 3}-\frac 1{\sqrt{2}} r_{\rho 3}\bigg)^2\nonumber\\
\xi^2_3(\theta)=&&\frac 23 r_{\lambda \perp}^2+\frac 23 {\rm cos}^2\theta r_{\lambda 3}^2
\eea
The full action (prior to analytical continuation) is (\ref{END1}) plus (\ref{NB2})
\bea
\label{FULLX}
S[\theta]\rightarrow \int_0^Td\tau 
&&\bigg(3em_Q^2+\frac 3{4e}\nonumber\\
&&+\frac 1{4e}(\dot{r_\lambda}^2+\dot{r_\rho}^2)
+\sigma_T \sum_{i=1}^3 |\xi_i(\theta)|\bigg)\nonumber\\
\eea

\subsection{Going to the light front frame}

For $\theta\rightarrow -i\chi$ and $T\rightarrow iT_M$, (\ref{FULLX})  analytically continues to the 
light front Hamiltonian or squared mass

\bea
\label{HLF}
H_{LF}=&&\sum_{i=1}^3\bigg(\frac{k^2_{i\perp}+m_Q^2}{x_i}\nonumber\\
&&+2\sigma_T \big(|i\partial /\partial x_i|^2+M^2r_{i\perp}^2\big)^{\frac 12}\bigg)
\eea
with the constraints: transverse $\sum_{i=1}^3k_{i\perp}=P_\perp=0$ and longitudinal $\sum_{i=1}^3 x_i=1$,  with the standard momentum fractions
$x_i=k_i^+/P^+$.

\subsection{A digression to 1+1 space-time}

The Hamiltonian derived above contains non-factorizable interaction between the longitudinal
and transverse coordinates which make the problem difficult. So, before we will address it in full, 
let us discuss  its longitudinal part  alone. The Hamiltonian
(\ref{HLF}) is then reduced to
\bea
\label{HLF4}
H_{LF,L}=\sum_{i=1}^3\bigg(\frac{m_Q^2}{x_i}
+2\sigma_T  |i\partial /\partial x_i|\bigg)
\eea
For a baryon in the star configuration,
(\ref{HLF4}) yields a longitudinal 
squared mass spectrum $M_n^2$, and parton amplitudes
$\varphi_n[x]$

\bea
\label{HLF5}
\sum_{i=1}^3\bigg(\frac{m_Q^2}{x_i}
+2\sigma_T |i\partial /\partial x_i|\bigg)\varphi_n[x]
=
M_n^2 \varphi_n[x]\nonumber\\
\eea
 Modulo the effective string tension  from the 3-dimensional reduction,
(\ref{HLF5}) is similar to the baryonic
 equation derived in 2-dimensional QCD~\cite{Bars:1976nk,Durgut:1976bc}.
 
 (\ref{HLF5}) can be regarded as the eigenvalue problem,
 for  3 identical particles with parton-x coordinates, moving  in
 a box  $0\leq x_i\leq 1$. If one naively  substitutes the potential by
 vanishing ( Dirichlet ) boundary condition $\varphi_n(x_i=0,1)=0$,
 the eigenstates are standing waves, e.g.
 \bea 
\label{SLATER}
 \varphi_n[x]\approx {2^{\frac 32}}\,
 \bigg({\rm sin}(n_1\pi x_1){\rm sin}(n_2\pi x_2){\rm sin}(n_3\pi x_3)\bigg)
\nonumber\\
\eea
with eigenvalues
\bea
\label{MASSPU}
M_n^2\approx 2\pi\sigma_T (|n_1|+|n_2|+|n_3|)
\eea
that reggeize along the diagonal $n_{1,2,3}=n\gg 1$ as

\bea
\label{SLOPESPURIOUS}
\alpha^\prime M_n^2\approx 3n
\eea
The factor of 3 reflects on the star configuration  with  three strings.

Unfortunately, this solution is very naive, for several reasons. The most obvious 
is that the  independent quantization of three quarks in a box,  ignores
the important momentum conservation constraint
 $$X=x_1+x_2+x_3=1$$ 
and therefore contains
spurious center of mass motion.
As already discussed in the previous section, one can use other coordinates which
are center of mass free. In particular, the Jacobi coordinates lead to a problem with
$two$ particles inside the equi-lateral triangle.

%
%
To solve this problem, we proceed in two steps. First, we unwind the 
square roots  by using the einbein trick once again 
\bea
\label{CONF2}
\sum_{i=1}^3\bigg|\frac{i\partial}{\partial x_i}\bigg|&=&\frac 12\bigg(\frac 1{e_{iL}}+e_{iL}\bigg(\frac{i\partial}{\partial x_i}\bigg)^2\bigg)\nonumber\\
&\rightarrow&\frac 12\bigg(\frac 3{e_{L}}+e_{L}\sum_{i=1}^3\bigg(\frac{i\partial}{\partial x_i}\bigg)^2\bigg)
\eea
and  assume equal $e_{iL}=e_L$ at the extrema, in the steepest descent approximation.
Second, we isolate  the center of mass coordinate, using Jacobi coordinates (\ref{eqn_Jacobi_x}).
%
The 3-particle laplacian in those coordinates is the sum of a  2-particle reduced Laplacian,
plus derivative of the center of mass variable

\bea
\label{SUMX}
\sum_{i=1}^3\bigg(\frac{i\partial}{\partial x_i}\bigg)^2=
\bigg(\frac{i\partial}{\partial\lambda}\bigg)^2+\bigg(\frac{i\partial}{\partial\rho}\bigg)^2+3 \bigg(\frac{i\partial}{\partial X}\bigg)^2
\nonumber\\
\eea
 For fixed center of mass $X=1$, (\ref{eqn_Jacobi_x}) maps the confining box-region $B=[0,1]^3$ for the coordinates $x_i$,
 to an equi-lateral triangle $\Sigma(x)$ of side $L=\sqrt{2}$, with
 corners located at 
 $$(\lambda, \rho)=\bigg(-\sqrt{\frac 23},0\bigg), \bigg(\frac{1}{\sqrt{6}}, \frac{1}{\sqrt{2}}\bigg), \bigg(\frac{1}{\sqrt{6}}, -\frac{1}{\sqrt{2}}\bigg)$$
The corners correspond to one particle carrying all the momentum, with the two others at rest.

The eigensystem of the first two terms in the Laplacian (now free from the center of mass motion!), amounts to solving
\bea
\label{REDX}
-\bigg(\frac{\partial^2}{\partial\lambda^2}+\frac{\partial}{\partial\rho^2}\bigg)
\varphi_{m_L,n_L}(\lambda, \rho)=e_{m_Ln_L}\varphi_{m_L,n_L}(\lambda,\rho)\nonumber\\
\eea
inside the triangle $\Sigma$, with Dirichlet boundary condition  $\varphi_{m_L,n_L}(\partial \Sigma)=0$.
Remarkably, although the solutions are not available for generic triangles,
  they are in fact known for equi-lateral triangles in closed form, found in \cite{RICHENS1981495}. Their existence
is due to the finite number of ray reflections, which make a closed set, 
as  explained in Appendix \ref{app_triangle}. The spectrum of the Laplacian is given by
\bea
\label{EMN}
e^D_{m_Ln_L}=\bigg(\frac{4\pi}{3L}\bigg)^2\bigg(\bigg(m_L-\frac {n_L}2\bigg)^2+\frac 34 n_L^2\bigg)\equiv \tilde{e}^D_{m_Ln_L}\pi^2\nonumber\\
\eea
with integer valued  longitudinal quantum numbers  $m_L,n_L$, restricted by $m_L\geq 2n_L$. The   states with $m_L>2n_L$ are doubly degenerate,
with normalized eigenstates~\cite{RICHENS1981495}
\bea
\label{BER1}
&&\varphi_{m,n}^{Dc}(\lambda, \rho)=\frac{4 }{L\,3^{\frac 34}}\bigg[{\rm cos}\bigg(\frac{2\pi(2m_L-n_L)\rho}{3L}\bigg)
{\rm sin}\bigg(\frac{2\pi n_L\tilde\lambda}{\sqrt{3}L}\bigg)\nonumber\\
&&-{\rm cos}\bigg(\frac{2\pi(2n_L-m_L)\rho}{3L}\bigg)
{\rm sin}\bigg(\frac{2\pi m_L\tilde\lambda}{\sqrt{3}L}\bigg)\nonumber\\
&&+{\rm cos}\bigg(\frac{2\pi(m_L+n_L)\rho}{3L}\bigg)
{\rm sin}\bigg(\frac{2\pi (m_L-n_L)\tilde\lambda}{\sqrt{3}L}\bigg)\bigg]\nonumber\\
&&\varphi_{m,n}^{Ds}(\lambda, \rho)=\frac{4 }{L\,3^{\frac 34}}\bigg[{\rm sin}\bigg(\frac{2\pi(2m_L-n_L)\rho}{3L}\bigg)
{\rm sin}\bigg(\frac{2\pi n_L\tilde\lambda}{\sqrt{3}L}\bigg)\nonumber\\
&&-{\rm sin}\bigg(\frac{2\pi(2n_L-m_L)\rho}{3L}\bigg)
{\rm sin}\bigg(\frac{2\pi m_L\tilde\lambda}{\sqrt{3}L}\bigg)\nonumber\\
&&-{\rm sin}\bigg(\frac{2\pi(m_L+n_L)\rho}{3L}\bigg)
{\rm sin}\bigg(\frac{2\pi (m_L-n_L)\tilde\lambda}{\sqrt{3}L}\bigg)\bigg]\nonumber\\
\eea
with $\tilde\lambda=\lambda+L/\sqrt{3}$. Their   symmetry properties include e.g. $\rho$ mirror symmetry
\bea
\varphi_{m_L,n_L}^{Dc,s} (\lambda, -\rho)=\pm \varphi^{Dc,s}_{m_L,n_L}(\lambda, \rho)
\eea
The Dirichlet  states with $m_L=2n_L$ are non-degenerate, with normalized eigenstates~\cite{RICHENS1981495}
\bea
\label{BER2}
&&\varphi^D_{2n_L, n_L}(\lambda, \rho)=\nonumber\\
&&\frac{2^{\frac 32}}{L\,3^{\frac 34}}\bigg[
2{\rm cos}\bigg(\frac{2\pi n_L\rho}{L}\bigg){\rm sin}\bigg(\frac{2\pi n_L\tilde\lambda}{\sqrt{3}L}\bigg)
-{\rm sin}\bigg(\frac{4\pi n_L\tilde\lambda}{\sqrt{3}L}\bigg)\bigg]\nonumber\\
\eea

Since (\ref{BER1}-\ref{BER2}) are separable in $(\lambda, \rho)$ and harmonic, they are readily seen to 
solve (\ref{SUMX}). The proof that these solutions form an orthonormal set on the triangle is
nontrivial, but we checked a number of cases explicitly.
Implicitly,  it follows from the 
observation that the mode number following from  (\ref{EMN}), saturates the so called Weyl  area rule~\cite{RICHENS1981495}.

\begin{figure}[htbp]
\begin{center}
\includegraphics[width=5cm]{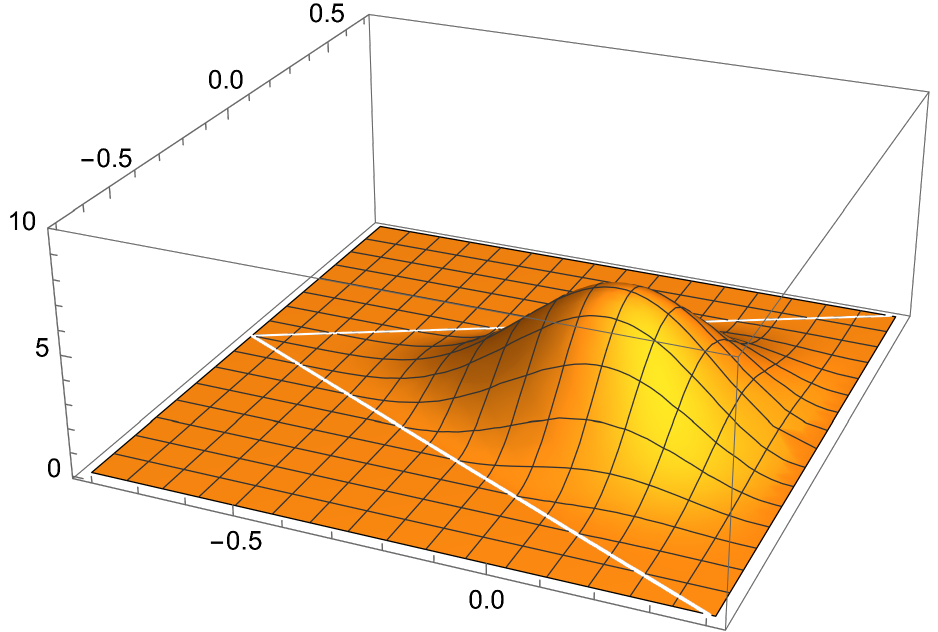}
\includegraphics[width=5cm]{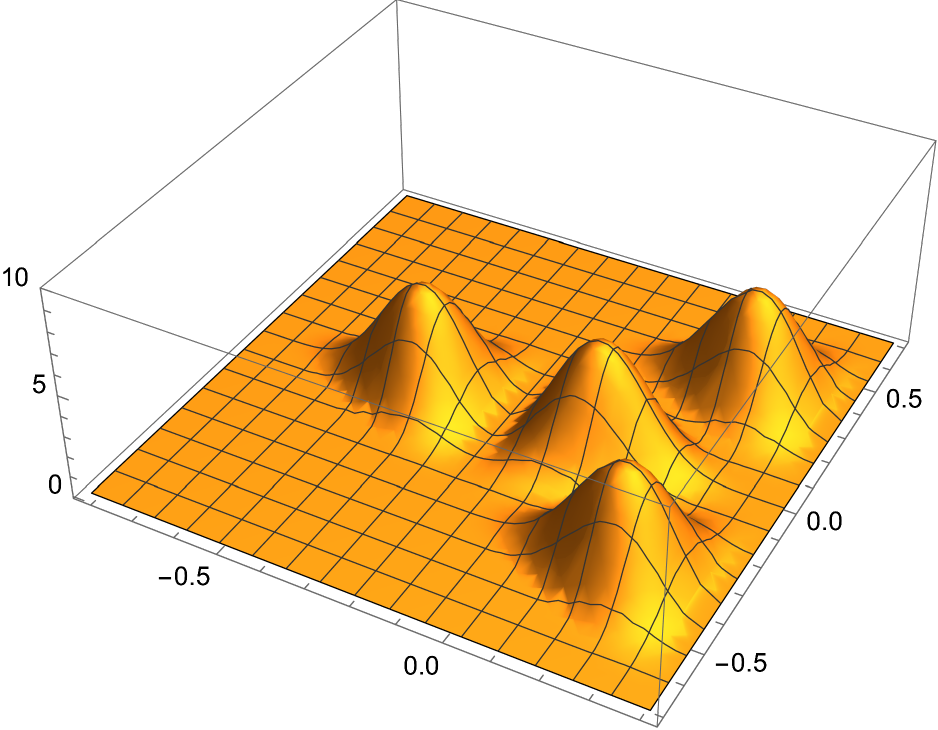}
\caption{Probability distribution $| \phi^D_{2n_L,n_L}|^2$ for $n_L=1$ (upper) and $n_L=2$ (lower) 
in the $\lambda,\rho$ plane, with manifest  mirror symmetry in $\rho$. }
\label{fig_state_N}
\end{center}
\end{figure}

We identify the   ground state  from the tower of states (\ref{BER2}) with
 $n_L=1$, and its radial excitations with   $n_L>1$. In Fig.~\ref{fig_state_N} we show the probability
 distributions for $n_L=1,2$.
 
These states are shown to Reggeize below.    We further note that  (\ref{BER2}), can be 
recast as  three standing waves with three ``momenta" $\tilde k$
\bea
\label{DSTANDING}
&&\varphi^D_{2n_L,n_L}(\lambda, \rho)=\nonumber\\
&&\frac{2^{\frac 72}}{L3^{\frac 34}}
{\rm sin}\bigg(\frac{2\pi n_L\tilde k_0}{\sqrt{3}L}\bigg){\rm sin}\bigg(\frac{\pi n_L\tilde k_+}{\sqrt{3}L}\bigg)
{\rm sin}\bigg(\frac{\pi n_L\tilde k_-}{\sqrt{3}L}\bigg)\nonumber\\
\eea
in the triangular domain limited by the sides
$$\tilde k_0=\sqrt{3}L/2\qquad \tilde k_\pm=\tilde\lambda\pm \sqrt{3}\rho=0$$
Remarkably, in  the original  x-Bjorken coordinates   the standing waves (\ref{DSTANDING}) are
identical to those in (\ref{SLATER}), for fixed   $X=x_1+x_2+x_3=1$, i.e.


\bea
\label{DSTANDINGX}
&&\varphi^D_{2n_L,n_L}(x_1,x_2,x_3)=(-1)^{n+1}\frac{2^{3}}{X3^{\frac 34}}\nonumber\\
&&\times
{\rm sin}\bigg(\frac{n_L\pi x_1}X\bigg){\rm sin}\bigg(\frac{n_L\pi x_2}X\bigg){\rm sin}\bigg(\frac{n_L\pi x_3}X\bigg)
\nonumber\\
\eea
(This observation perhaps allows for the extension of the Dirichlet standing states  and their  excitations,
to the states of  more exotic hadrons with $N>2$ compact 
multi-quark content  -- tetraquarks, pentaquarks, hexaquarks
\bea
\varphi_{n_L}^D(x_1, ..., x_N)=\frac{C_N}{X}\prod_{i=1}^N{\rm sin}\bigg(\frac{n_L\pi x_i}{X}\bigg)
\eea
with $X=\sum_{i=1}^Nx_i$, and the normalization  $C_N$ fixed by the polygonal  volume, set
 by the longitudinal momentum constraint $X=1$. The meson case with $N=2$, of course requires a single 
standing wave, as we used in our previous papers.)

Using (\ref{EMN}), the  contribution of the Laplacian to  the baryon  spectrum  is 
\bea
\label{RADFREE}
\Delta M_{m_Ln_L}^2\approx&& 2\pi\sigma_T\sqrt{3\tilde{e}^D_{m_Ln_L}}\nonumber\\
\approx&& \bigg(\frac{4}{\sqrt{6}}\bigg)\bigg(2\pi\sigma_T\bigg(\bigg(m_L-\frac {n_L}2\bigg)^2+\frac 34 n_L^2\bigg)^{\frac 12}\bigg)\nonumber\\
\eea
For large  quantum numbers,  it  reggeizes into a linear dependence. For
 the ground state and its radial excitations series, with $m=2n$ in (\ref{BER2}), its contribution to the spectrum can be
 compared to the conventional Regge trajectory of the mesons $n=\alpha' M^2$ with $\alpha^\prime=1/2\pi\sigma_T$
\bea
\label{M2NN}
\alpha^\prime \Delta M_{2n_L,n_L}^2\approx 2\sqrt{2}\, n_L
\eea
There is an additional factor of $2\sqrt{2}\approx 2.83$, to be also compared with  the
Regge slope from the spurious spectrum  (\ref{SLOPESPURIOUS}), where this factor is just 3, the number of strings.
\\
\\
Let us add the following consideration. Although the ``cup" potential $V=\sum m_Q^2/x_i$ was not yet
included, its very existence -- especially for heavy masses $m^2_Q/2\sigma_T\gg 1 $ --
motivated us to look at standing waves that  vanish at the cup's boundaries. 
In the opposite limit of light quarks $m^2_Q/\sigma_T\leq 1$  we can ignore
the confining potential   as a small perturbation, and thus ignore its
end-point constraint. (Recall that $m_Q$ is not the ``Lagrangian" quark mass but an effective
one, including the constituent quark mass. So  even for light quarks $m_Q\sim 350\, MeV$,
 while $ \sqrt{\sigma_T}\approx 400\, MeV$, so this limit may be of  academic interest only.)

In this case, it is perhaps more appropriate to use  free end-point 
or $Neumann$ boundary conditions 
$$\varphi^\prime_{n_L}(x_i=0,1)=0$$ so as to minimize the {\it kinetic}
contribution for the excited states. Again, 
the  eigenstates free of center of mass motion, can be sought using also the  ray reflection method, as we
suggest in  Appendix~\ref{app_triangle}. 

In particular, the Neumann analogue of the tower
of Dirichlet standing states (\ref{DSTANDING}) built on the ground state,  is
readily found as
\bea
\label{NSTANDING}
&&\varphi^N_{2n_L,n_L}(\lambda, \rho)=\nonumber\\
&&\frac{2^{\frac 72}}{L3^{\frac 34}}
{\rm cos}\bigg(\frac{2\pi n_L\tilde\lambda}{\sqrt{3}L}\bigg){\rm cos}\bigg(\frac{\pi n_L\tilde\lambda_+}{\sqrt{3}L}\bigg)
{\rm cos}\bigg(\frac{\pi n_L\tilde\lambda_-}{\sqrt{3}L}\bigg)\nonumber\\
\eea
or equivalently in x-Bjorken 
\bea
\label{NSTANDINGX}
&&\varphi^N_{2n_L,n_L}(x_1,x_2,x_3)=(-1)^{n_L}\frac{2^{3}}{X3^{\frac 34}}\nonumber\\
&&\times
{\rm cos}\bigg(\frac{n_L\pi x_1}X\bigg){\rm cos}\bigg(\frac{n_L\pi x_2}X\bigg){\rm cos}\bigg(\frac{n_L\pi x_3}X\bigg)
\nonumber\\
\eea
with $X=1$ subsumed.
The solutions (\ref{NSTANDING})  satisfy Neumann boundary conditions in the triangular domain by inspection,
with the same spectrum and Regge trajectory as Dirichlet for $m_L=2n_L$, i.e. $e_{2n_L,n_L}^N=e_{2n_L,n_L}^D$
but  with $n_L=0$ a priori allowed.


 \subsection{Spectrum of the diagonal part of the Hamiltonian $H_0$}~\label{sec_H0}
 
Our  light front Hamiltonian contains three parts
$$H_{LF}\approx H_{0\perp}+ H_{0 x_i} +\tilde V_{LF}$$
The first two are transverse oscillator and longitudinal ``triangular cup": for
both of them we managed to find their complete set of eigenfunctions. The  remaining residual part
is not amenable to analytic integration,  and will be treated numerically.
Also 
we  use the einbein trick to get rid of the square roots in the confining term
\bea
\label{HLF2}
H_{LF}\approx &&\sum_{i=1}^3\bigg(\frac{k^2_{i\perp}+m_Q^2}{x_i}\nonumber\\
&&+\sigma_T \bigg(3a +\frac 1a \sum_{i=1}^3(|i\partial/\partial x_i|^2+(3m_Q)^2b_{i\perp}^2\big)\bigg)\bigg)\nonumber\\
\eea
with $M\approx 3m_Q$ used on the right-hand-side to close the mass squared operator.
Again,  we assumed  equal einbeins 
$a_i\rightarrow a$ in (\ref{HLF2}) by steepest descent. To the first kinetic term we add and subtract its value
at $x_i=\frac 13$, producing an oscillator with fixed frequency, and a residual potential
$\tilde V$ which is close to zero at the center
of the triangular cup. 

In terms of the Jacobi coordinates, the diagonalizable part  reads
\bea
\label{HLF3}
&&H_{0LF}=3(\vec p_\rho^2+\vec p_\lambda^2+3m_Q^2) \\
&&+\frac {\sigma_T}a
\bigg(|i\partial/\partial\lambda|^2+|i\partial/\partial\rho|^2+(3m_Q)^2(\vec b_\lambda^2+\vec b_\rho^2)\bigg)\nonumber
\eea
where the all the vectors are in the  transverse plane, and $\vec b_\lambda, \vec b_\rho$ are coordinates conjugate
to the corresponding momenta. To elucidate the dependence on $a$ we rewrite it as
\bea
\label{HLF4}
&&M^2_0(n_\lambda, n_\rho, n_L, m_L)=(3m_Q)^2\nonumber\\
&&+\frac{\sigma_T}{\sqrt{a}}M_\perp^2(n_\lambda, n_\rho)+\frac{\sigma_T}a
M_L^2(m_L, n_L)+3\sigma_T a\nonumber\\
\eea
with
\bea
&&M_L^2(n_L, m_L)=e^D_{n_L, m_L}\nonumber\\
&&M_\perp^2(n_\lambda, n_\rho)=\frac{6\sqrt{3}m_Q}{\sqrt{\sigma_T}}(n_\lambda+n_\rho+2)
\eea
The einbein in (\ref{HLF4}) minimizes the squared mass, and is solution to
the quartic Ferrari equation
$$6 \sqrt{a}^4-M_\perp^2\sqrt{a}-2M_L^2=0$$
For large longitudinal quantum numbers $n_L, m_L\gg 1$ the squared mass
reggeizes $$M^2_0\approx 2\sqrt{3}\sigma_T M_L$$ as we noted earlier. However, for large transverse quantum numbers
$n_\lambda, n_\rho\gg 1$ the squared mass does not 
\bea
M_0^2\approx 18\sigma_T\bigg(\frac{M_\perp^2}{6\sigma_T}\bigg)^{\frac 23}
\eea

\begin{figure}[h]
\begin{center}
\includegraphics[width=6cm]{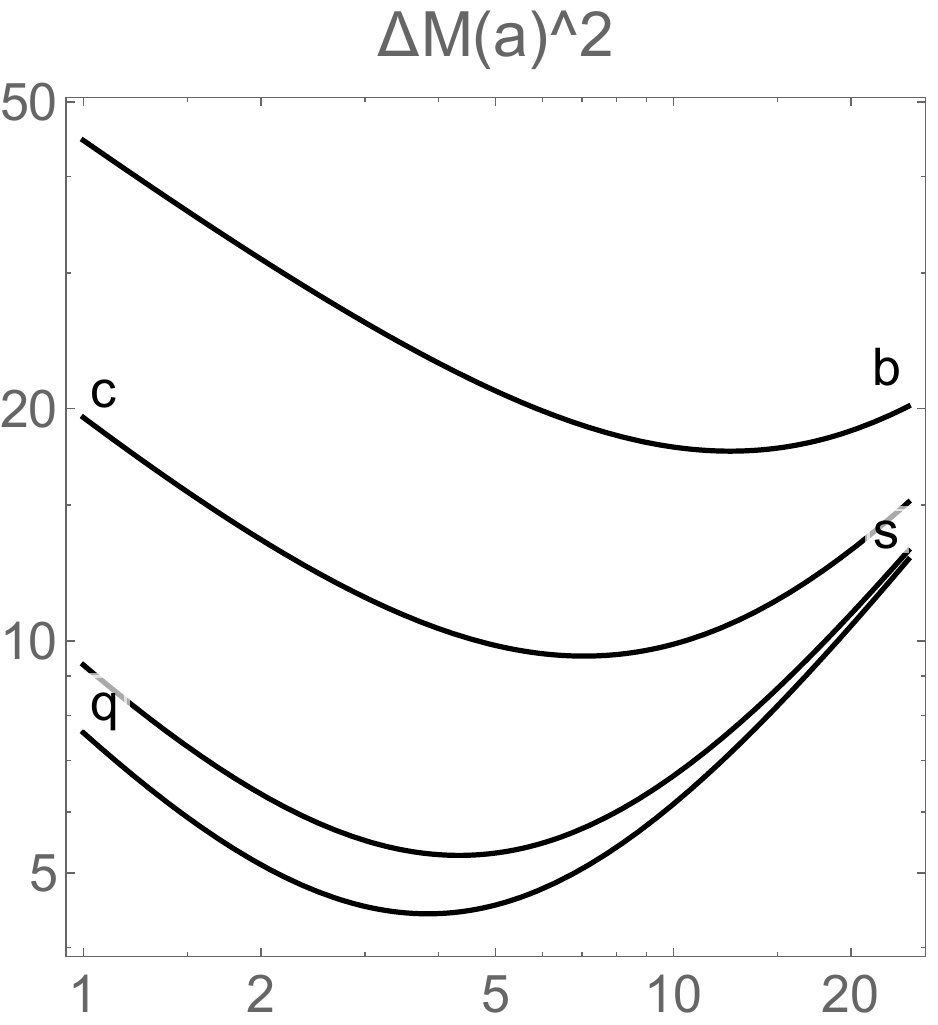}
\caption{The lowest eigenvalue of $H_{0LF}$ in $GeV^2$ versus the (dimensionless) ``einbine parameter" $a$, for $b,c,s,q$ quarks. Using this plot we perform the minimization in $a$. }
\label{fig_M2_of_a}
\end{center}
\end{figure}

Recall that the results following from $H_{0LF}$ are still to be modified by the additional residual contributions,
 stemming from  $\tilde V_{LF}$ to be added below, but which are  independent
of our variational parameter $a$. Therefore the minimization over $a$ can already be performed numerically. 
With our standard values for the string tension $\sigma_T=(0.4\, GeV)^2$,  and quark masses $b,c,s,q$,
 we show in Fig.~\ref{fig_M2_of_a} the dependence on $a$ of the lowest eigenvalue for each species.
 
\section{The non-factorizable potential $\tilde V$}
The non-factorizable part of the potential is
\ba \tilde V&=&{\vec p_1^2+m_Q^2\over x_1} + {\vec p_2^2 +m_Q^2\over x_2} \nonumber\\
&+& {\vec p_3^2+m_Q^2\over x_3} - 3 (\vec p_1^2 +\vec p_2^2 + \vec p_3^2)-9m_Q^2 
 \ea
Using the  Jacobi coordinates for the transverse and longitudinal momenta, we get
\ba  \label{eqn_tilV}
\tilde V=&& -\bigg((3 (-2 p_\lambda p_\rho (\sqrt{6} - 6 \lambda) \rho + 
       9 m_Q^2 (2 \lambda^2 + \sqrt{6} \lambda^3 \nonumber\\ 
     &&\qquad+ 2 \rho^2 - 
          3 \sqrt{6} \lambda \rho^2) + 
       p_\lambda^2 (9 \lambda^2 + 3 \sqrt{6} \lambda^3 + 
          3 \rho^2  \nonumber \\ 
       &&\qquad+ \sqrt{6} \lambda (1 - 9 \rho^2)) + 
       p_\rho^2 (3 \lambda^2 + 3 \sqrt{6} \lambda^3 + 9 \rho^2 \nonumber \\
       &&\qquad-
           \sqrt{6} \lambda (1 + 9 \rho^2))\big) \bigg)\nonumber\\
           &&\qquad\times\bigg( { 1 \over -2 + 
     9 \lambda^2 + 3 \sqrt{6} \lambda^3 + 9 \rho^2 - 
     9 \sqrt{6} \lambda \rho^2}\bigg)\nonumber\\
     \ea
For zero orbital motion,  the two oscillators are independent, and  the  term $\langle p_\lambda p_\rho \rangle $ vanishes on  average. 
$\langle p_\lambda^2\rangle$ and $ \langle  p_\rho^2 \rangle $ are directly related to the number of quanta $n_\lambda,n_\rho$, and
so one  has to calculate only the matrix entries in terms of all possible longitudinal quantum numbers
 $\langle n_L,m_L | \tilde V | n_L' m_L' \rangle$, see Appendix~\ref{sec_tilV}.
    
    \subsection{Masses of the states} 
With the evaluation of the matrix   $\tilde V$ and its eigenvalues,
our  technical task is completed. We now
 can finally  carry  the calculation of the
full eigenvalues -- squared masses of  the flavor symmetric baryons, for the four quark flavors $b,c,s,q$. 
We keep here longitudinal quantum numbers to their lowest values $n_L=1,m_L=2$,
and assume that the  transverse oscillators are excited as a function of a single  $n=(n_\rho+n_\lambda)/2 $.

\begin{figure}[h]
\begin{center}
\includegraphics[width=7cm]{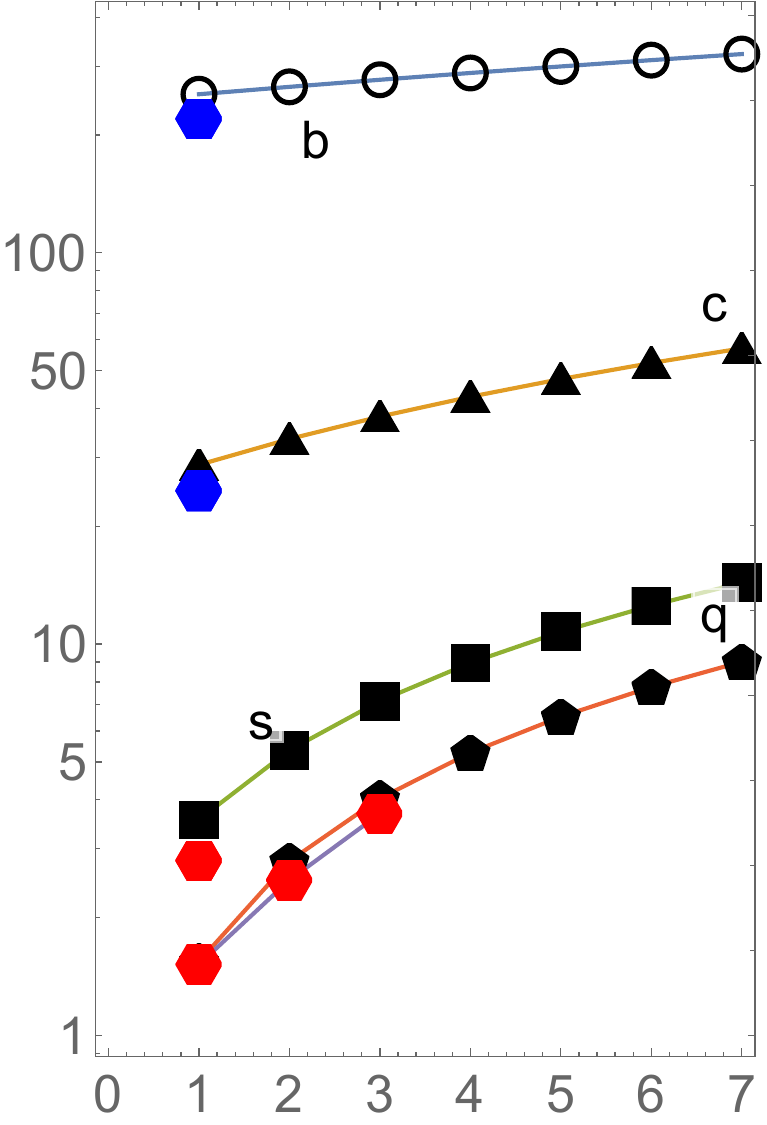}
\caption{Squared masses of baryons $M_{n+1}^2(Q,\frac 32)$ in $GeV^2$,  versus the principal quantum number $n+1=1..7$. The black  circles, triangles, squared and pentagons are results of our calculations for the  flavors  $b,c,s,q$. The red hexagons  are the experimental values of three $\Delta^{++}$ and one $\Omega^-$ masses, from PDG. 
The two blue  hexagons are  model predictions for masses of $ccc$ and $bbb$ baryons, from Table I.}
\label{fig_all_masses}
\end{center}
\end{figure}

Our results for squared masses are shown  in black-symbols in Fig.~\ref{fig_all_masses}. For comparison, we show the experimental masses in red hexagons. The
  blue hexagons are available model predictions for  $ccc$ and $bbb$ baryons. Since a constant in LF Hamiltonian
 remains undefined, we fixed one constant for all masses, so that the mass of the $uuu$ baryon $\Delta^{++} $ is set
 to experimental one.
 For definiteness, the plots corresponds to effective masses of $u,s,c,b$ quarks to be $0.28,0.45,1.5,4.8\, GeV$, respectively, not specially fitted to this plot but inherited from meson studies.

Finally, let us remind that all the calculations were done for traditional ``star" or $Y$ model of confinement.
Transition to (perhaps more accurate)  ansatz A leads to the same Hamiltonian with a string tension rescaling
(\ref{RESCALING}), downwards by about 13\% like one  has observed it in static potentials.

 \section{Wave functions of the states}

 Our main results are not the masses of the $\frac 32^+$ states and their radial excitations, 
 but their corresponding wave functions. The ground states
 in all channels, have a transverse momentum dependence that is about Gaussian,
 e.g. $\psi(p_\perp) \sim exp(-\beta^2 p_\perp^2/2)$. The scale 
parameter $\beta$ is related to the mass and frequency of the effective oscillator $\mu,\omega$,
 \be \beta= \sqrt{\mu \omega} = \bigg({a  \over 3 m_Q^2 \sigma_T} \bigg)^{\frac 14} 
  \ee   
The mean square of the transverse momentum is  approximately
\be \langle p_\perp^2 \rangle \approx \beta^{-2} \approx 0.942, 0.466, 0.183, 0.104 \, (GeV^2)    \ee
for the $b,c,s,q$ $3/2^+$ single-flavor baryons.   

The longitudinal wave functions for  heavy  quark masses $m_Q$, are defined mainly by the $O(m_Q^2)$ 
part of the potential $\tilde V$. They are  discussed in Appendix~\ref{app_triangle}, and illustrated in 
Fig.~\ref{fig_tilV_wf} by the solid-black line. In general, since the Hamiltonian is $H_0$ plus $\tilde V$, the wave functions lie
 between their eigenfunctions, or between the solid and dashed line in Fig.~\ref{fig_tilV_wf}. We note that the
difference is mostly around the negative maximun of $\lambda$, or when $x_3$ is close to 1.
 Since the triangle is equi-lateral and
the wave function is symmetric, this implies that the
suppression in fact occurs near all three corners of the triangle.

\section{ Longitudinal   wave functions, for two-string or V configuration} 
Finally,  we now show how the  baryon configuration  with a  ``quark-on-junction", or two-string V configuration,
once reduced to longitudinal part, can also be solved. In this configuration, 
the central quark can be e.g. a heavy one, or it can be $d$ quark jumping between
two $u$s (reminiscent to the  1-folded string~\cite{Bardeen:1975gx}).
In this case it is more convenient to use another set of coordinates
\bea
\label{PARTONX123}
x_1=\frac 13 + y_1\qquad x_2=\frac 13 +y_2-y_1\qquad x_3=\frac 13-y_2\nonumber\\
\eea
which are also canonically conjugate $[y_i, r_i]=i\delta_{ij}$. In those variables the physical domain is an
isosceles triangle with corners at (2/3, 1/3), (-1/3, -2/3), (-1/3, 1/3), or
 half the   2-dimensional square.

With this in mind, we 
can recast (\ref{HLF5}) as
\bea
\label{MN3}
M_n^2\varphi_n[y]&=&\bigg(\frac {m_1^2}{\frac 13+y_1}+\frac {m_2^2}{\frac 13+y_2-y_1}+\frac {m_3^2}{\frac 13-y_2} \nonumber \\
&+&2\sigma_T(|i\partial_{y_1}|+|i\partial_{y_2}|)\bigg)\varphi_n[y]
\eea
Using again  the einbein trick to make the confining part quadratic, and changing the momenta to derivatives over coordinates,
yields a similar eigenvalue problem as that for the star configuration. The only difference is that the cup potential is now an 
isosceles triangle instead of an equilateral one. 
Remarkably, Laplacian problem on it also admits an exact solution, by the eikonal ray construction same as
for the equilateral case. The solutions 
are the superposition of the symmetric and antisymmetric   standing waves on the unit square
 
 \bea
 \label{STANDSOL}
 &&\varphi^{\pm}_{m_L,n_L}(x_1,x_3)=\nonumber\\
 &&\sqrt{2}\big({\rm sin}(m_L\pi x_3){\rm sin}(n_L\pi x_1)\pm {\rm sin}(n_L\pi x_3){\rm sin}(m_L\pi x_1)\big)\nonumber\\
 \eea
 with $y_{1,2}$ re-expressed in terms of parton   $x_{1,3}$ given in (\ref{PARTONX123}),
and  the corresponding eigenvalues
 \bea
 \label{SPEC}
&& e^+_{m_L,n_L}=\pi^2(m_L^2+n_L^2)\qquad m_L=n_L\pm 1, n_L\pm 3, ..\qquad\nonumber\\
 && e^-_{m_L,n_L}=\pi^2(m_L^2+n_L^2)\qquad m_L=n_L\pm 2, n_L\pm 4, ..\qquad\nonumber\\
 \eea
 (\ref{STANDSOL}) are symmetric or anti-symmetric under the exchange of $x_{1,3}$ (exchange of the u-quarks at the end points),
 and both vanish for parton-x $x_{1,3}=0$ and $x_1+x_3=1$ or equivalently $x_2=0$, as expected. The lowest state corresponds\
 to $m_L=n_L+1=2$, with a PDF for the u-quark
 \bea
  \label{LOWEST}
 u(x_1)=\int_0^{1-x_1} dx_3 |\varphi^{+}_{21}(x_1,x_3)|^2 \rightarrow \frac{(2\pi^2)^2}{3} (1-x_1)^5\nonumber\\
 \eea
which is again very soft at large parton-x.

 The V-string quantum spectrum follows  the same reasoning as that for the star baryon, with the
 result
\bea
\label{MN2}
\alpha^\prime M_{m_\pm}^2=(m_+^2+m_-^2)^{\frac 12}
\eea
where  $m_\pm = m_L\pm n_L$.
(\ref{MN2}) Reggeizes with the meson slope of $1$  along $m_-$. These
Regge trajectories are the quantum states in correspondence with the  classical yo-yo states,
noted in~\cite{Bardeen:1975gx}.

\begin{figure}[htbp]
\begin{center}
\includegraphics[width=6cm]{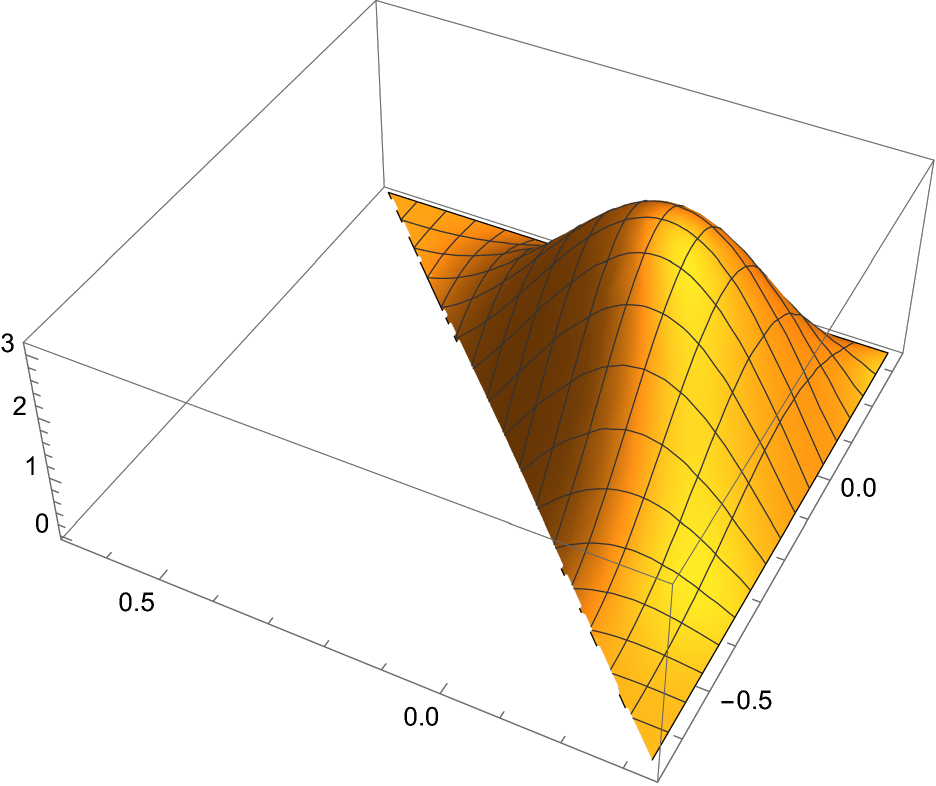}
\includegraphics[width=6cm]{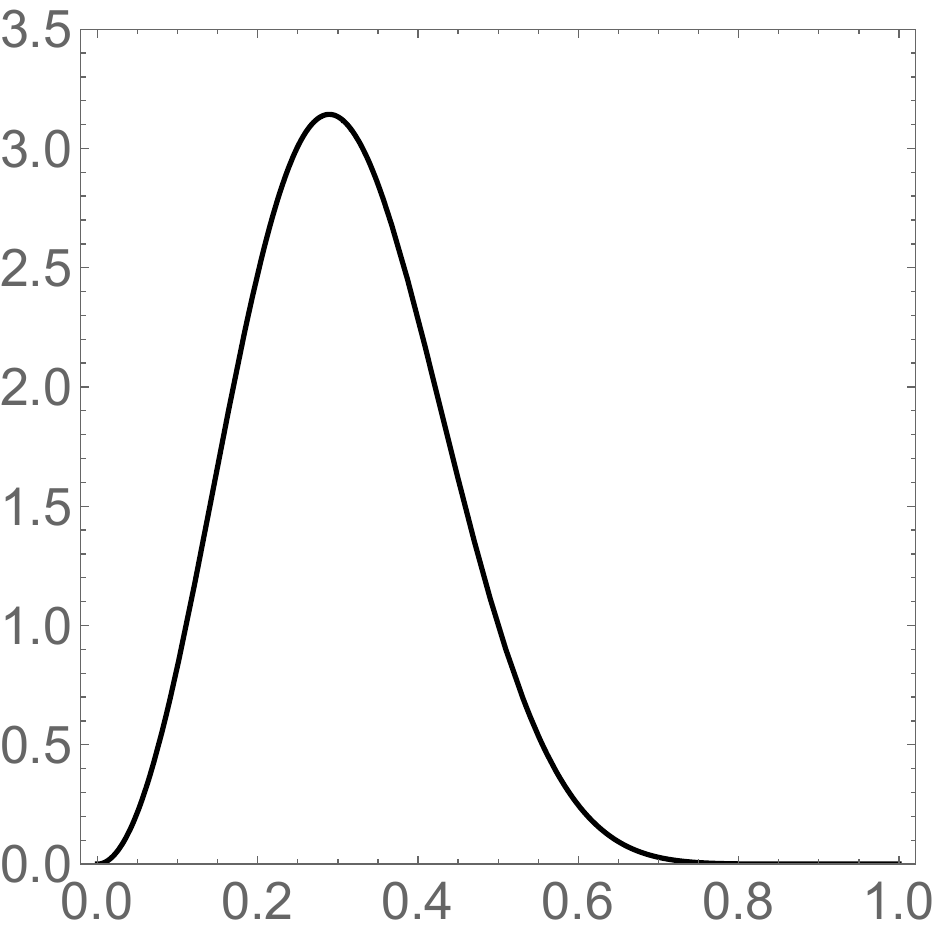}
\caption{Upper: ground state wavefunction  $\varphi_{21}^+(x_1,x_3)$ of the V-baryon in (\ref{STANDSOL}). Lower: 
Corresponding PDF for the ground state in (\ref{LOWEST}).
}
\label{fig_q_on_j_wf}
\end{center}
\end{figure}

With the inclusion of  the ``cup potential"  induced by the kinetic energy we also solved the problem numerically.
The lowest state on this triangle is shown in Fig.\ref{fig_q_on_j_wf}(upper). In a standard way we also calculated DAs
and PDFs for quarks: due to lack of symmetry, they are different for those in $45^o$ and $90^o$ corner quarks.
The PDF for the former is shown in  Fig.\ref{fig_q_on_j_wf}(lower): and as one can see it also
is way too ``soft" at $x\rightarrow 1$ as compared to observed PDF of the nucleon.

\section{Conclusions}

We  start by recalling  the  chief goals of this series of papers, namely 
to bring together the QCD studies of hadronic spectroscopy and light front observables -- DAs, PDFS, GPDs etc. 
The use of the light front formulation brings in certain issues, but enforces  relativistic kinematics, 
for both the light and heavy quarks,  and allows the study of their bound state structure  in a single framework.

We believe that hadronic physics should have the same logical and methodical
structure, as that in atomic and nuclear physics. Quantum dynamics should be defined by a Hamiltonian,
which can be simplified at first but focus on the main physics. For example, in this paper
we have included the  kinetic term, and the confining term  only, leaving the more complicated contributions 
(Coulomb and spin-dependent forces) for next studies. 

Once the Hamiltonian is defined, its diagonalization yields physical states, which automatically possess
such  basic properties as mutual orthogonality.  
 This part of the problem is rather
technical and took most of this paper. We have shown that 
one can do so, either by using diagonalization in certain basis set, or numerically solve for wave functions
(as we did for longitudinal ones). 
Any wave function is complete, in the sense that one can calculate any observable from it, say various PDFs, GPDs
and formfactors by standard general formulae. Of course, those are defined  at normalization scale
with {\it low resolution}, at which gluons and antiquarks are suppressed. Comparison with data at high
resolution scale of experiments should include their evolution by well developed perturbative QCD tools.

In our attempts
to $derive$ the light-front  Hamiltonian $H_{LF}$  we have focused on the nonperturbative interquark interaction.  
We started by relating the lattice data to few simple models of confinement, Y, V and A models. 
They follow from
 relativistic QCD strings (flux tubes) with a  linear dependence on the size of the baryon. Then we 
 performed our own evaluation of  the three-quark potentials from Wilson line correlators, based  on a semiclassical 
 model of the Euclidean QCD vacuum using a  ``dense instanton liquid".  We  find a reasonably good agreement
 between the results and lattice numerical data. We  also see that out of three empirical models, the Ansatz A
 seems to be the closest to these potentials. 
  
 The important distinctions between our approach and other versions of $H_{LF}$ 
in the literature are among others: \\(i) $H_{LF}$  is derived
from  established QCD lattice facts; \\(ii) We do not rush to fit any parameters to the experimental data but use only
  standard values for the effective quark masses, and the string tension $\sigma_T$;\\ 
 (iii) We do not even start with the nucleon, for which obviously there is large set of experimental data,
 but study single-flavor baryons because they should not have strong diquark clustering which is known to be
 flavor-antisymmetric. \\ (iv) We do not focus on only the ground state but
 consider several excited  states in each quark channel, so that the dependence on principal and orbital
 quantum numbers can be available.  It is imperative that the Hamiltonian used do have reasonable
predictions for many states, at least they approximately reproduce the expected Regge behavior.

This paper is mostly technical in nature, in it we showed {\em how one can solve the quantum mechanical problem} of three
 identical quarks, given their interaction by a confining string. Unlike other approaches in the light front literature, we 
  use as many degrees of freedom as needed: six  Jacobi coordinates/momenta for three-quark baryons. There is no need 
    to ``subtract spurious center of mass motion".  We do quantum mechanics in momentum representation, and therefore,
 in the longitudinal direction we show how to solve a Schroedinger-like
 equation on physical triangle. The ``unfactorizable" potential is defined, set into its minimal form and 
 its effect on the wave functions is evaluated,
 by diagonalization in the approriate basis or by numerical solution for subset of degrees of freedom.  

This paper focuses on flavor symmetric  ($QQQ$) baryons with maximal spin $\frac 32$,
 the most symmetric setting
 to start with it. Perturbative Coulomb forces -- belived to be important for heavy quarks -- are not yet included,
 as well as spin-dependent forces.
Effects related to t'Hooft instanton-induced Lagrangian -- believed to be important for light quarks and
contributing to ``diquark"  correlations  -- will be discussed in the sequels of this series.

\vskip 1cm
{\bf Acknowledgements}

This work is supported by the Office of Science, U.S. Department of Energy under Contract No. DE-FG-88ER40388.

\appendix

\begin{widetext}
\section{Averaging over the $SU(3)$ rotations} \label{sec_Weingarten}
The color averaging over the unitary matrices can be carried using the Weingarten coefficients
\bea
\label{WEI}
&&\bigg<U^{a_1}_{c_1}U^{\dagger b_1}_{d_1}
U^{a_2}_{c_2}U^{\dagger b_2}_{d_2}
U^{a_3}_{c_3}U^{\dagger b_3}_{d_3}\bigg>_U=\nonumber\\
+&&\frac{(N_c^2-2)}{N_c(N_c^2-1)(N_c^2-4)}\sum_{n=1}^{3!}\delta^{a_1a_2a_3}_{(d_1d_2d_3)_n}\delta^{c_1c_2c_3}_{(b_1b_2b_3)_n}\nonumber\\
-&&\frac{1}{(N_c^2-1)(N_c^2-4)}\sum_{n=1}^{3!}\delta^{a_1a_2a_3}_{(d_1d_2d_3)_n}
\bigg(\delta^{c_1c_2c_3}_{(b_2b_1b_3)_n}+\delta^{c_1c_2c_3}_{(b_1b_3b_2)_n}+\delta^{c_1c_2c_3}_{(b_3b_2b_1)_n}\bigg)\nonumber\\
+&&\frac{2}{N_c(N_c^2-1)(N_c^2-4)}\sum_{n=1}^{3!}\delta^{a_1a_2a_3}_{(d_1d_2d_3)_n}
\bigg(\delta^{c_1c_2c_3}_{(b_3b_1b_2)_n}+\delta^{c_1c_2c_3}_{(b_2b_3b_1)_n}\bigg)
\eea
where the  short-hand notation used
\bea
\delta^{a_1a_2a_3}_{(d_1d_2d_3)_n}\equiv \sum^{3!}_{n=1}\bigg(\delta^{a_1}_{d_1}\delta^{a_2}_{d_2}\delta^{a_3}_{d_3}+{\rm perm.}\bigg)
\eea 
refers to the product of three kroneckers in the sum over the n! permutations of the permutation group $S_3$. Inserting (\ref{WEI}) into (\ref{3WP1}), 
and using the identities in~\cite{Shuryak:2021iqu}, we have
\bea
\label{3WP2}
&&+\bigg[\bigg(\frac{2}{N_c}\bigg)^3\,c_1c_2c_3\bigg] \delta^{a_1}_{b_1}\delta^{a_2}_{b_2}\delta^{a_3}_{b_3}\nonumber\\
&&+\bigg[\frac{1}{2N_c(N_c^2-1)}\,c_1\,{\rm Tr}(-is_2\tau\cdot n_2 \lambda_2^A){\rm Tr}(-is_3\tau\cdot n_3 \lambda_3^A)\bigg] \delta^{a_1}_{b_1}(\lambda_2^B)^{a_2}_{b_2}(\lambda_3^B)^{a_3}_{b_3}+{\rm 2\, perm.}\nonumber\\
&&+\bigg[\frac{N_c}{8(N_c^2-1)(N_c^2-4)}\,d^{IJK}\,{\rm Tr}(-is_1\tau\cdot n_1 \lambda_1^I){\rm Tr}(-is_2\tau\cdot n_2 \lambda_2^J){\rm Tr}(-is_3\tau\cdot n_3 \lambda_3^K)\bigg] d^{ABC}\,(\lambda_1^A)^{a_1}_{b_1}(\lambda_2^B)^{a_2}_{b_2}(\lambda_3^C)^{a_3}_{b_3}\nonumber\\
&&+\bigg[\frac{8}{8N_c(N_c^2-1)}\,f^{IJK}\,{\rm Tr}(-is_1\tau\cdot n_1 \lambda_1^I){\rm Tr}(-is_2\tau\cdot n_2 \lambda_2^J){\rm Tr}(-is_3\tau\cdot n_3 \lambda_3^K)\bigg] f^{ABC}\,(\lambda_1^A)^{a_1}_{b_1}(\lambda_2^B)^{a_2}_{b_2}(\lambda_3^C)^{a_3}_{b_3}\nonumber\\
\eea
which is explicit in terms of the invariants of $SU(3)_c$. 
Furthermore,
the $SU(2)_c$ color structure of the instanton is not enough to support the last two terms with $f,d$ structure constants in (\ref{3WP2}). 
\end{widetext}

\section{The ``einbine" trick} \label{sec_einbine}
For completeness, we briefly recall the use of the einbein trick.  Consider an  expression (operator $X$) of the 
form
\be X \rightarrow {1 \over 2} \bigg(a X^2+{1\over a}\bigg) \ee
with a variational parameter $a$.  The  r.h.s. can be readily used in a Hamiltonian. Noting that the r.h.s. has a minimum as a
function of $a$, and that its value at the minimum is the l.h.s., the latter follows from the former.

This trick is used in section \ref{sec_semiclassics}  to eliminate  the  square root of the relativistic kinetic energy
 $\sqrt{p^2+m^2}$ in the CM frame, and in the rest of the paper by changing from linear confinement to its
quadratic form.

\section{Laplacian operator  and basis functions for $s,t$  map (\ref{eqn_long_map})} \label{sec_app_st}
Given a coordinate mapping $y^m=(s,t,u)$, there is a well defined procedure for the re-writing of the Laplacian 
in differential geometry, using  the metric tensor  $g$, 
\begin{widetext}
\label{GMN}
\ba g = \begin{gmatrix}[b]
1/8 (u^2 + 2 t u^2 + t^2 u^2) && 1/8 (s u^2 + s t u^2) && 1/8 (s u + 2 s t u + s t^2 u) \\
   1/8 (s u^2 + s t u^2) && 1/8 (3 u^2 + s^2 u^2) && 
   1/8 (-u + s^2 u + 3 t u + s^2 t u)\\
   1/ 8 (s u + 2 s t u + s t^2 u) && 1/8 (-u + s^2 u + 3 t u + s^2 t u) && 
   1/8 (3 + s^2 - 2 t + 2 s^2 t + 3 t^2 + s^2 t^2) 
    \end{gmatrix}
\ea
\end{widetext}
The covariant Laplacian is  given by the expression
\be L={1\over \sqrt{g}} {\partial \over \partial y^m} \sqrt{g} g^{m n} {\partial \over \partial y^n} \ee
where $g^{m n}$ is the inverse matrix to $g_{m n}$ in (\ref{GMN}), and $g=det(g_{m n})$. Only the determinant is
simple enough to be quoted here
$$ g={u^4 \over 64} (1+t)^2 $$
The full expression for the Laplacian is involved, and will not be given here. Fortunately one of the variables --
the ``scale" $u$ -- can be set  to one, and all derivatives over it neglected. Then the result takes the somehow simple form
\bea  &&L[\phi]={1 \over (1 + t)^2} \big(+2 (1 + t)^2 (-1 + 3 t) 
{\partial \phi(s,t) \over \partial t} \nonumber \\
&&+ (3 +4t+ 2 t^2+4 t^3+ 3 t^4) 
{\partial^2 \phi(s,t) \over \partial t^2} 
+ 16 s 
{\partial \phi(s,t) \over \partial s} \nonumber \\
&& +(- 8 s + 8 s t^2 ) {\partial^2 \phi(s,t) \over \partial t \partial s}
  + (8 + 8 s^2){\partial^2 \phi(s,t) \over \partial s^2} \big)\nonumber\\
\eea

The physical wave functions must have boundary conditions that are consistent with 
finiteness of the Laplacian. (And the ``nonfactorizable" potential $\tilde V$,
which is proportional to $1/(x_1 x_2 x_3)$ and thus singular at the boundaries.)
Therefore  all  wave functions can be supplied with some suppression factors, which for this map may take
 a factorized  form 
$$ \Psi(s,t)=(1-s^2)  (1-t)^a (1+t)^b \Phi(s,t)$$
with some parameters $a,b$ ensuring finiteness of all matrix elements.
The remaining non-singular function $\Phi(s,t)$  can be  conveniently expressed as
products of 
Jacobi polynomials $P_n^{A,B}(s) P_{n^\prime}^{C,D}(t)$, with indices $A,B,C,D$ related to $a,b$
and the invariant measure
$\sqrt{g}$. 

\section{Basis functions for transverse momenta} \label{sec_basis}
The LFWFs part  depends on  two (Jacobi) transverse momenta  (\ref{eqn_Jacobi}) $\vec \rho_\perp,\vec \lambda_\perp$, via a double set of 2d Harmonic oscillator wave functions. Those are the same
as the ones we used in our previous papers for mesons. They can be written in compact form
using generalized Laguerre polynomials 
\ba  \psi_{n,m}(p_\perp,\beta) &=&{\sqrt{2} \over 2\pi} \beta \sqrt{n! \over (n+|m|)!} e^{-p_\perp^2\beta^2/2+i m \phi}  \nonumber \\
&\times& (p_\perp  \beta)^{|m|}L_n^{|m|}\big( \beta^2 p_\perp^2\big)
\ea
where $\vec p_\perp$ stands for  $\vec \rho$ or $\vec\lambda$.
The value of beta comes from $H_0$ as discussed in section~\ref{sec_H0}.

\begin{figure}[htbp]
\begin{center}
\includegraphics[width=8cm]{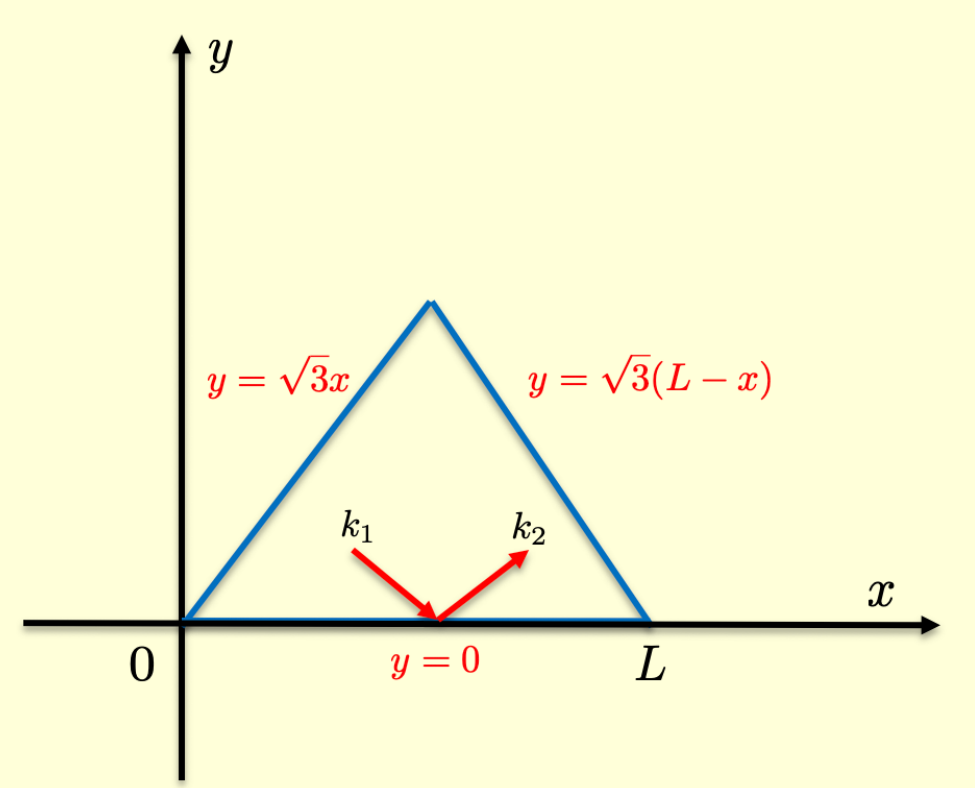}
\caption{Equilateral triangle used for the light-ray construction, with the $k_{1,2}$ rays shown. See text.} 
\label{fig_triangle}
\end{center}
\end{figure}

\section{Confined quantum waves in an equilateral triangle}~\label{app_triangle}

In this Appendix, we outline the derivation of the confined quantum waves by an equi-lateral triangle,
which leads to (\ref{BER1}-\ref{BER2}), as solutions to the 2-dimensional Shroedinger equation~(\ref{SUMX}),
with the result given in~\cite{RICHENS1981495}.
In the absence of the confining boundary conditions, the solutions are separable in harmonic waves $e^{\pm ik\cdot r}$,
with $\vec k=(k_x, k_y)$ a 2-dimensional real wave-vector in the $(x, y)$ coordination used in~\cite{RICHENS1981495} (or
$x=\rho+L/2$ and $y=-\lambda+L/2\sqrt{3}$ in our coordination) as illustrated in Fig.~\ref{fig_triangle}. 
The triangle confines the waves, by enforcing the wave amplitudes to vanish at the triangle edges. We now
explain how.

Following
the construction outlined in~~\cite{RICHENS1981495}, consider the equi-lateral triangle, with corners
$(x,y)=(0,0), (0,L), (L/2, \sqrt{3}L/2)$ in the labeling  of~\cite{RICHENS1981495}
(Our triangular domain in $(\lambda, \rho)$ follow by the mapping quoted above). A  ray in the equilateral triangle with arbitrary wave-vector
$\vec k$, can undergo six distinct reflections before emerging parallel to itself 
(arbitrary triangles would not close under reflection). Each pair of rays at the triangle
side fulfills the null amplitude condition. As a result, we can pair the rays at each side,  to generate the solution to
(\ref{SUMX}) with triangular confining walls, by superposition thanks to the linearity of the Sturm-Liouville problem.
6 ray reflections, map the equilateral triangular phase space, onto that of a torus phase space, and close the reflection cycle.
We note that half the equilateral triangle, and the right-angle isosceles triangle (half of a  square), 
are two other triangular domains, with a closed reflection phase space as well. (Incidentally, these are also the
triangles, for which the Coulomb problem using the image  construction,  closes with a finite number of images.)

Let $\vec k_{i=1,..,6}$ be the 6 
reflected wave-vectors, initiated by the  harmonic solution with $\vec k_1=(k_x,k_y)$,
 that close the reflection cycle in the equi-lateral triangle. The reflection of $\vec k_1$ by the
wall at $y=0$, generates the reflected wave-vector  $\vec k_2=(k_x, -k_y)$. By construction,
the  harmonic combination
\bea
\label{ONE}
\big(e^{-ik_1\cdot r}-e^{-ik_2\cdot r}\big)(x, y=0)=0
\eea
solves (\ref{SUMX}) since $k_1^2=k_2^2$, and satisfies the null condition on the triangle side  $y=0$. The 
next pair of reflections on the triangle side  $y=\sqrt{3} x$, yields the pair of wave-vectors
$\vec k_3=-\frac 12(k_x+\sqrt{3}k_y,-\sqrt{3}k_x+k_y)$ 
and 
$\vec k_4=-\frac 12(k_x-\sqrt{3}k_y,-\sqrt{3}k_x-k_y)$. 
The superposition (\ref{ONE}) with the substitution $1,2\rightarrow  3,4$,
again solves (\ref{SUMX}) since $k_3^2=k_4^2$, and satisfies the null condition on the triangle side  $y=\sqrt{3}x$. The same
reasoning applies to the last pair of reflections back at the wall at $y=0$, with 
$\vec k_5=-\frac 12(k_x-\sqrt{3}k_y,\sqrt{3}k_x+k_y)$ 
and $\vec k_6=-\frac 12(k_x+\sqrt{3}k_y,\sqrt{3}k_x-k_y)$, followed by the substitution $1,2\rightarrow 5,6$ in (\ref{ONE}).
The quantization of the wave-vectors in (\ref{BER1}-\ref{BER2}) with Dirichlet boundaries,
follow by the null condition of

\bea
\label{TWO}
&&\varphi^D_k(x,\sqrt{3}(L-x))=\nonumber\\
&&\bigg(\sum^6_{i=1} (-1)^{i+1}e^{-ik_i\cdot r}\bigg)(x,\sqrt{3}(L-x))=0
\eea
on the last triangular side  $y=\sqrt{3}(L-x)$. Note that all wave contributions
carry $\pm 1$ weight by reflection symmetry. This is the solution derived  in~\cite{RICHENS1981495},
and used in our analysis, after pertinent rotation and translation of the equi-lateral triangle quoted here.

Finally, we note that the Neumann boundary conditions on the equi-lateral triangle,  can be
enforced using a similar reasoning  to that for Dirichlet, but with all ray contributions carrying the
same sign,  and (\ref{TWO}) changed to

\bea
\label{THREE}
&&\varphi^N_k(x,\sqrt{3}(L-x))=\nonumber\\
&&\bigg(\sum^6_{i=1} e^{-ik_i\cdot r}\bigg)(x,\sqrt{3}(L-x))=0
\eea

\begin{figure}[htbp]
\begin{center}
\includegraphics[width=6cm]{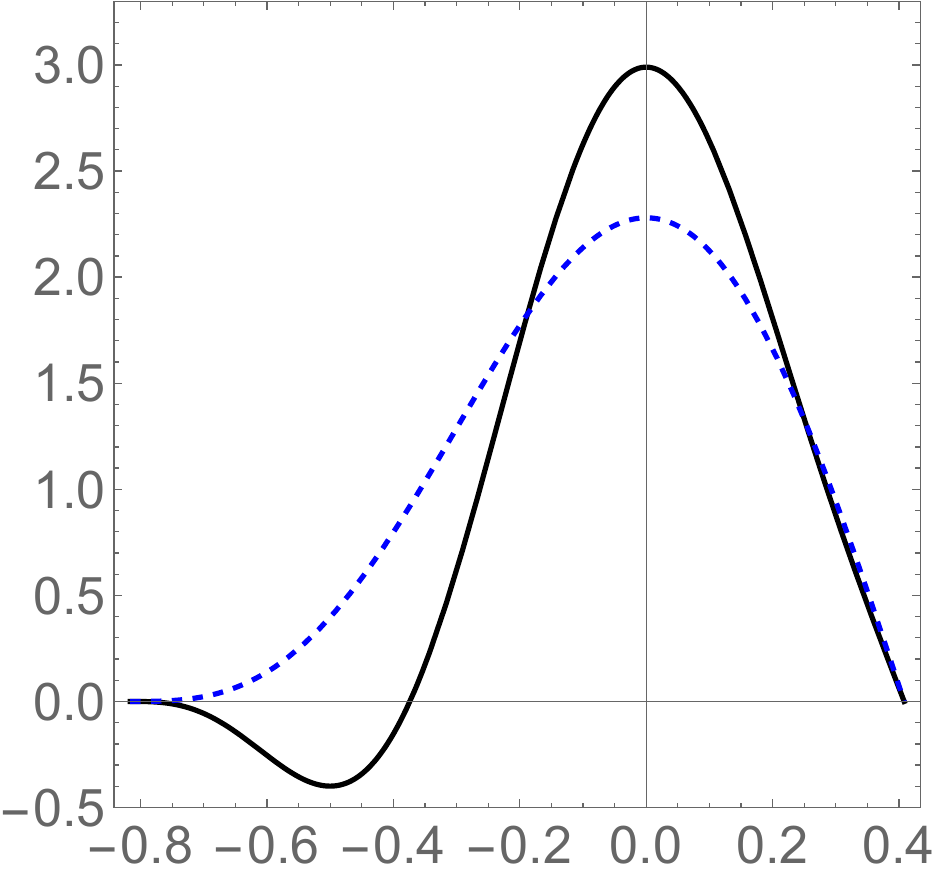}
\caption{The eigenfunction~(\ref{eqn_tilV_wf}) at $\rho=0$ as a function of $\lambda$ is shown by the  solid-black curve, to be compared to the lowest standing wave (eigenstate of $H_0 $ for $n_L=1$)
as a dashed-blue curve (\ref{BER2}) .}
\label{fig_tilV_wf}
\end{center}
\end{figure}

\section{The non-factorizable potential $\tilde V$} \label{sec_tilV}
The expression of this potential in Jacobi coordinates has been given in (\ref{eqn_tilV}),
and here we discuss its matrix elements. For that we use the family of states (\ref{BER2}), with a single quantum number
$n_L$.

For definiteness, we only give the matrix elements proportional to $p_\lambda^2, p_\rho^2$ (the ones proportional to $m_Q^2$ are similar). 
Specifically, the upper $4\times 4$ part is
\begin{widetext}
\ba
{\langle n_{L1} | \tilde V | n_{L2} \rangle  \over p_\lambda^2}=
\begin{pmatrix} 
 2.26289  &1.90816 & 1.31215 & 0.991796 \\ 
1.90816 & 5.75718 & 2.73486 & 2.04179 \\
 1.31215 & 2.73486 & 7.95585 & 3.30115 \\
0.991796 & 2.04179 & 3.30115 & 9.56256 
\end{pmatrix}  \nonumber \ea
\end{widetext}
Note that it grows strongly with increasing quantum number, and then it decreases away from the diagonal,
although rather slowly. Therefore the eigenstates involve strong mixing, and are not very close to
the original states  (\ref{BER2})  (which are eigenstates of the Laplacian in  the triangle).
In particular, the eigenvalues of this matrix are
$    1.42047 , 4.14098, 5.98336,13.9937 $, somewhat spreading from the diagonal matrix elements. 
Only the lowest state eigenvector (in the original basis) shows
dominance of the lowest states, with strong decrease of the higher state admixture
$C_l=0.920434, -0.390426, -0.017908, -0.00695135$, hence the  probabilities $|C_l^2|$ for $n_L>2$ are
negligibly small. Note also that the corrections  are all negative. To explain what these corrections do, we 
show the eigenfunction 
\be \label{eqn_tilV_wf}
\psi_{\tilde V}=\sum_l C_l \varphi^D_{l, l}(\lambda, \rho) \ee
as a solid-black curve in Fig.~\ref{fig_tilV_wf}. The dashed-blue curve is the exact and lowest standing wave 
in (\ref{BER2}) for $n_L=1$. 

Since the Hamiltonian is a combination of $H_0$ (thus the Laplacian) and $\tilde V$, physical wave functions
for the ground state baryons must be {\em in between} the two curves shown in Fig.~\ref{fig_tilV_wf}.

\bibliography{baryons,mesons-at-CM1,mesons-at-CM}
\end{document}